\RequirePackage{fix-cm}
\documentclass[smallextended]{svjour3}       
\usepackage{graphicx}
\usepackage{amssymb}
\usepackage{amsmath}
\usepackage{nomencl}
\usepackage{color}
\usepackage{graphicx}
\usepackage{epstopdf}
\usepackage{mathptmx}      
\usepackage{float}
\usepackage{enumerate}
\usepackage{booktabs}

\usepackage{url}
\usepackage{mathtools}
\usepackage{bm}
\usepackage{esvect}
\usepackage{appendix}
\usepackage{color}
\usepackage{multirow}
\usepackage[authoryear,numbers]{natbib}
\usepackage{threeparttable}
\usepackage{booktabs}
\usepackage{verbatim}
\usepackage{makecell}
\usepackage{rotating}
\usepackage{diagbox}

\makenomenclature

\smartqed  
\usepackage{graphicx}
\begin{document}

\title{A single-camera synthetic Schlieren method for the measurement of free liquid surfaces 
}

\titlerunning{A single-camera method for the measuring of liquid surfaces}        

\author{Huixin Li, Marc Avila and Duo Xu    
}


\institute{Huixin Li \at
              University of Bremen, Center of Applied Space Technology and Microgravity (ZARM), 28359 Bremen, Germany \\
            \and
            Marc Avila \at
            University of Bremen, Center of Applied Space Technology and Microgravity (ZARM), 28359 Bremen, Germany \\
 			\and
			Duo Xu \at
			The State Key Laboratory of Nonlinear Mechanics, Institute of Mechanics, Chinese Academy of Sciences, 100190 Beijing, China\\
			School of Engineering Science, University of Chinese Academy of Sciences, 100049 Beijing, China\\
			University of Bremen, Center of Applied Space Technology and Microgravity (ZARM), 28359 Bremen, Germany \\
			\email{duo.xu@imech.ac.cn} (Corresponding author)           \\
 \\  
}

\date{Received: date / Accepted: date}

\maketitle

\begin{abstract}
A single-camera synthetic Schlieren method is introduced to measure the height of a dynamic free liquid surface in a container with a flat bottom. Markers placed on the bottom, seen through the free surface, are virtually displaced due to light refraction at the surface. According to Snell's law, the marker displacements depend on the refractive indices of the transparent liquid and the air, and on the surface height and its spatial gradients. 
We solve the resulting governing nonlinear equation with the Newton--Raphson method to obtain the surface height. Our method does not require a reference surface height, which allows the measurement of surfaces topography in situations where the liquid depth is unknown. We demonstrate the accuracy of the method by performing experiments of surface ripples and dam-break flows, and discuss the measurement uncertainty, errors and limitations. 
\end{abstract}

\section{Introduction}
\label{intro}
The {quantitative} measurement of a liquid free surface is fundamentally important in many applications. For example, in marine and coastal engineering, wind-generated surface waves have substantial impacts on marine vessels and offshore platforms \citep{Holthuijsen2007}. In chemical engineering, falling liquid films with dynamic free surface are employed for the efficient heat exchange in steam condensers \citep{Karimi1998}. The liquid sloshing with dynamic free surface motions are essential in propellant tank in aerospace devices and in liquid natural gas cargos in ship industry \citep{Ibrahim2005}.
{Quantitative} optical imaging measurements of the height of the free surface are advantageous because of their non-intrusive nature. 

One imaging methodology associates the height of the free surface with light absorption in the liquid. Given that image magnification is independent of surface height, \citet{jahne2005combined} used a telecentric optical system {to measure the surface height, which is a function of }the light absorption rates from light-emitting diodes (LEDs) with red and near-infrared lights. {Here a challenging task is to precisely align the collimated beams in different wavelengths.}
{\citet{aureli2014combined} used a camera with two charged-coupled devices (CCDs) including a monochrome near-infrared sensor in order to implement the surface measurements with the working principle of light absorption. They took the advantage of high absorption capacity of water in the near-infrared range. Their method can reach an accuracy in the order of $1$~mm, however the measurement sensitivity decreases with the water depth.}
 
Planar laser-induced fluorescence (LIF) with a laser sheet has also been used to capture the profiles of the free surface \citep{Duncan99,Andre14}. A fast-scanning of a laser sheet can be used to obtain and reconstruct the surface topography in two dimensions {\citep{van_meerkerk2020scanning}. The scanning needs to be fast enough so that the surface structure can be approximately assumed to be frozen. In the LIF method, edge-detection algorithms are required to extract the surface profiles from the LIF images}, and the physical length corresponding to an image pixel determines the measurement resolution. The measurement dynamic range (the ratio of the maximum measured value to the smallest measured value) is determined approximately by the image resolution.

Light reflection can also be used for surface measurements. For example, the slope of two-dimensional surfaces can be measured with a polarimeter by correlating the surface orientation with changes of light polarization \citep{zappa2008retrieval}. {This technique can be used not only in laboratory but also in rivers with unpolarized skylight or moon-glade \citep{zappa2008retrieval,vinnichenko2020measurements}. Alternatively, symbol patterns and fringe patterns are projected onto the  surface, and the deformation of the patterns due to the surface topography is exploited for the surface measurements \citep{tsubaki2005stereoscopic, cobelli2009global, hu2015quantification}. The intensity of the light reflection is crucial to the measurement accuracy, and additives mixed with the liquid (water) are employed to improve the light reflection at the surface.  }

A group of imaging techniques relies on light refraction. \citet{kurata1990water} used a camera to capture a reference and a refraction image of a striped grating, which was placed below a  shallow water channel. When the water surface had ripples, the imaged grating was virtually displaced with respect to the still, horizontal water surface, due to the changed light refraction. {In their method, the topography of the free surface is obtained from the virtual displacements. The surface height can also be obtained provided that the height of a reference surface point is known. In their flow, there is a point that the surface height is nearly unchanged in time, so that they use the height at this point as the reference height to reconstruct the surface depth from the topography data.}
\citet{fouras2008measurement}, \citet{moisy2009synthetic} and \citet{ng2011experimental} further developed this method by replacing the grating with random dot patterns, so that optimized cross-correlation particle-image-velocimetry (PIV) algorithms can be used to compute the virtual displacements precisely. Alternatively to the dot patterns, \citet{zhang1994measuring} introduced a color-coded pattern, where colors were designed to correlate with surface slopes one-to-one \citep{dabiri1997}.  

The advantages of refraction methods are the use of low-cost illumination, simple optical configuration and a single camera. Despite their simplicity and precision, {a drawback of these methods is that a reference height is needed to reconstruct the height of the surface from the surface topography.} In some situations, the liquid depth is not known and the method cannot be applied. 
\citet{morris2004image} and \citet{gomit2013free} used two or more cameras to obtain stereoscopic views of the same pattern. They could directly obtain the topography and the height of the free surface by matching the surface orientation at each surface point from the measurements of multiple cameras \citep{morris2004image} (see table~\ref{tab:ref}). However, using multiple cameras raises the hardware costs, results in more time-consuming calibration of the camera positions, and requires more computation power in the post-processing. Recently, \citet{engelen2018spatio} proposed a method  
with a single camera for the surface measurements, which relies on the condition that the surface topography can be depicted by a parametric model. 

We present an extension of the free-surface synthetic Schlieren (FS-SS) method of \citet{moisy2009synthetic}, which allows to directly measure the topography and the height of the free liquid surface using a single camera, without neither reference height nor parametric models. In \S\ref{sec:principle}, the working principle is introduced. Two sets of experiments were carried out to test our method. Experiments of surface ripples, where the spatially averaged height is approximately constant, are shown in \S\ref{sec:ripples}. Experiments of dam-break flows, where the averaged height changes in time, are presented in \S\ref{sec:dam_break}. A detailed discussion and conclusions are given in \S\ref{sec:discussion} and~\S\ref{sec:conclusion}, respectively.

\begin{table}
	\centering
	\renewcommand{\arraystretch}{1.1}
		\begin{tabular}{p{0.15\textwidth}  p{0.1\textwidth} p{0.12\textwidth}  p{0.12\textwidth}  p{0.1\textwidth} p{0.2\textwidth} }
			\toprule
			Reference & Quantity of camera & Method & Object \& pattern & Illumination & Measurability of liquid height \\ \midrule
			\cite{zhang1994measuring} & 1 & Refraction & Random colored blocks & Halogen light & {Yes, with reference height}  \\
			\cite{dabiri1997} & 1 & Refraction / Reflection & Random colored blocks & -- &  {Yes, with reference height}  \\
			\cite{morris2004image} & 2 & Refraction \& pattern & Checkerboard & LED & Yes \\
			\cite{jahne2005combined} & 1& Refraction \& light absorption & Gridded dots & LED & Yes, with complex telecentric illumination  \\
			\cite{fouras2008measurement} & 1 & Refraction \& Image correlation & Random dots & LED & Yes, with reference height \\
			\cite{moisy2009synthetic} & 1 & Refraction \& Image correlation & Random dots & LED & Yes, with reference height   \\
			\cite{gomit2013free} & 3 & Refraction & Random PIV tracers & Laser sheet &Yes  \\
			\cite{aureli2014combined} & 1 & Refraction \& light absorption & Random colored blocks & LED & Yes, with coupled RGB Bayer array and Monochrome sensor \\
			\cite{engelen2018spatio} & 1 & Refraction \& pattern match & Checkerboard & LED & Yes, with parametric surface model \\
			\cite{kolaas2018bichromatic} & 2 & Refraction \& image correlation & Random dots &  bichromatic LED & Yes, with reference height and telecentric lens \\
			\bottomrule
		\end{tabular}
		\caption{A brief list of refraction methods for free surface measurements.}\label{tab:ref}
\end{table}

\section{Methodology}
\label{sec:principle}
\begin{figure}[htb]
	\centering
	\includegraphics[width=0.6
	\textwidth, trim={0cm 0cm 0cm 0cm}, clip]{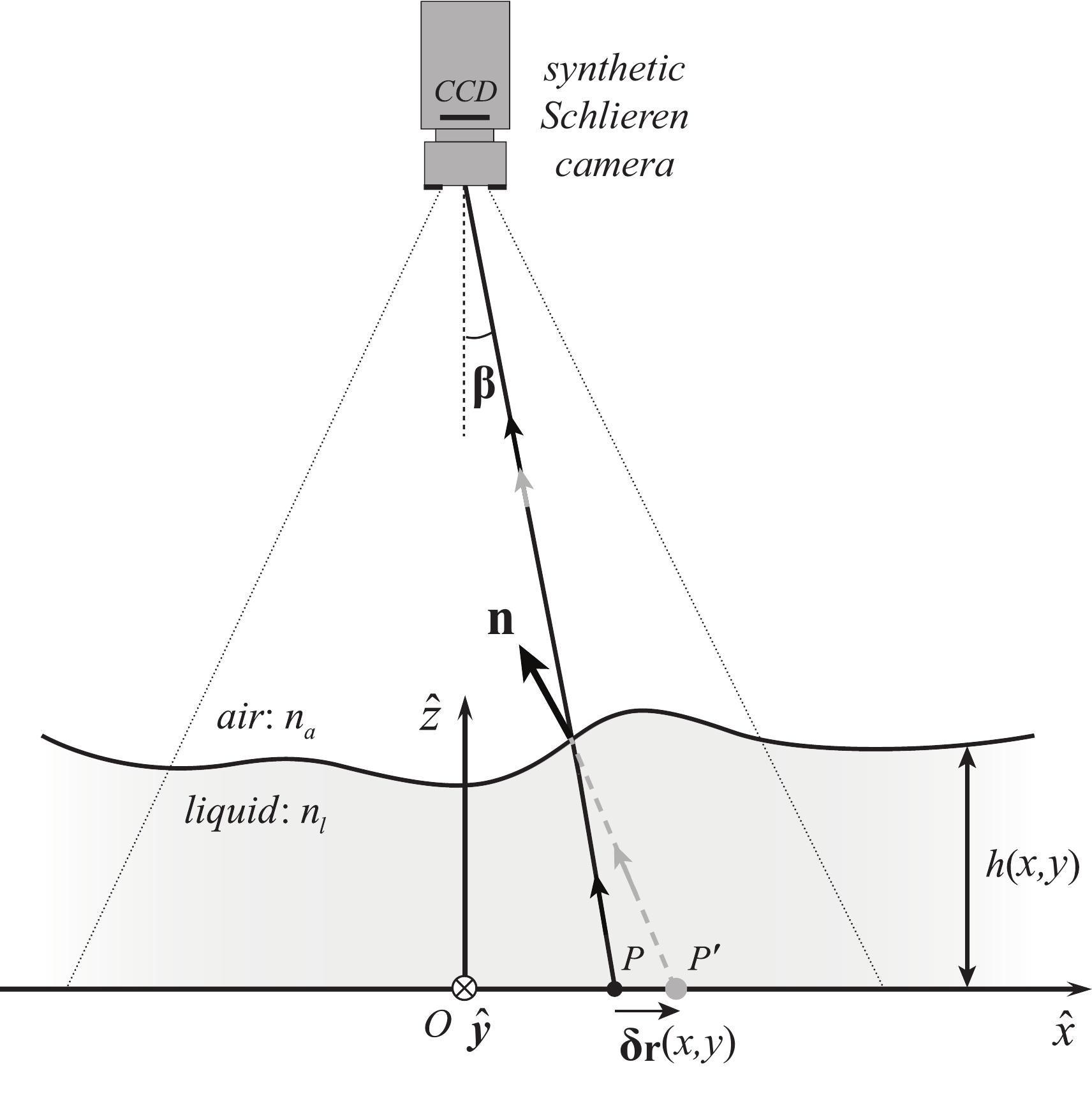}
	\caption{\label{fig:method} Schematic of working principle of measuring the topography and the height of free air-liquid surface $h(x,y)$ with a surface normal direction $\textbf{n}$. $(\hat{x}, \hat{y}, \hat{z})$ and $O$ denote the Cartesian coordinate and the origin, respectively. The displacement from a point $P$ to the point $P'$ (due to the surface refraction) is $\delta \textbf{r}$. $\bm{\beta}$ is the angle between the light (black line with an arrow) and the camera axis.	}
\end{figure}
The presented methodology exploits the light refraction at the air-liquid interface. A schematic of the working principle, based on the synthetic Schlieren method \citep{moisy2009synthetic} and the background-oriented Schlieren method \citep{Raffel2015BOS},  is shown in figure~\ref{fig:method}. When there is no liquid, the incident light ray at a point $P$ follows a straight trajectory (shown in black) toward the camera. When there is a liquid-air interface, the incident light ray at point $P'$ follows the dashed gray trajectory toward the surface and then follows the same black trajectory after leaving the surface. The light refraction is governed by the Snell's law for refractive index of the liquid $n_l$ and that of the gas $n_a$ (in this study $n_l=1.334$ and $n_a=1$, corresponding to water and air). The displacement $\delta \textbf{r}$ from the point $P$ to the point $P'$ depends on the liquid height $h(x, y)$, and its spatial gradients $\nabla h(x, y)$, according to the equation 
\begin{equation}\label{eq:full}
\delta \textbf{r} = h \left[ \mathrm{tan}\left(\mathrm{tan^{-1}}(\nabla h) + \mathrm{sin^{-1}}[(n_a/n_l) \cdot \mathrm{sin}(\bm{\beta}-\mathrm{tan^{-1}}(\nabla h))]\right) - \tan(\bm{\beta}) \right],
\end{equation}
{where $\bm{\beta}=\beta_x(x,y)\hat{x}+\beta_y(x,y)\hat{y}$ denotes the angle of the camera viewing the point $P$. The detailed derivation of this equation is given in Appendix.}
When the camera is set to satisfy the paraxial approximation, $\bm{\beta}(x,y)$ is small so that $\tan(\bm{\beta})\approx \bm{\beta}$. Similarly, $\tan(\nabla h) \approx \nabla h$ when $\nabla h$ is small, e.g. $\nabla h<0.5$, see \S\ref{subsec:simp_full}. 
{Then equation~\eqref{eq:full} can be simplified to}
\begin{equation}\label{eq:simp}
	\delta \textbf{r} = h \cdot (1-n_a/n_l) \left(\nabla h - \bm{\beta} \right),
\end{equation}
where $h(x,y)$ is the only unknown to be determined, provided that $\delta \textbf{r}$ and $\bm{\beta}$ have been pre-computed, as explained in what follows.
\begin{figure}[htb]
	\centering
	\includegraphics[width=0.7\textwidth, trim={0cm 0cm 0cm 0cm}, clip]{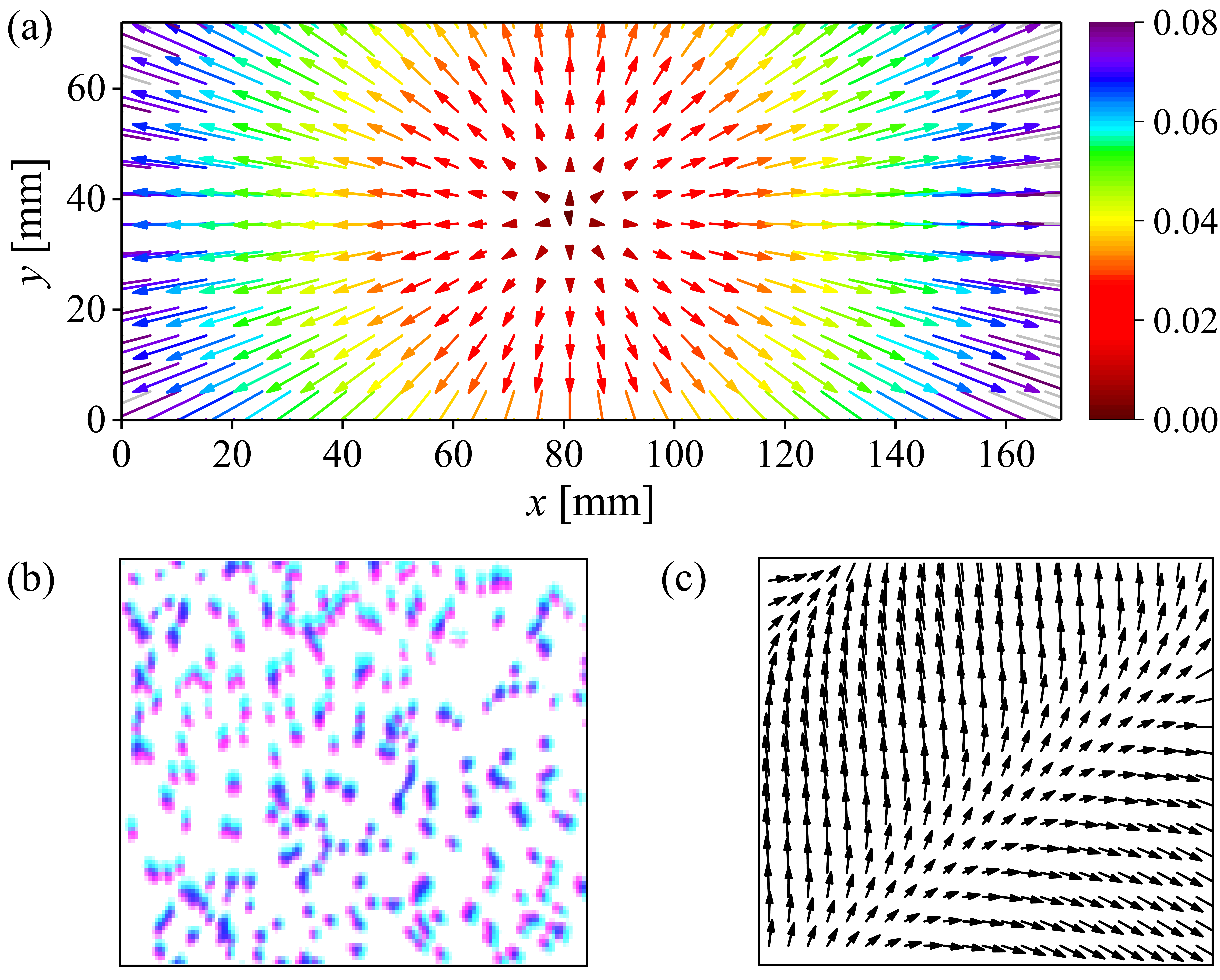}
	\caption{(Color online) (a) The vector map of $\bm\beta(x,y)$ obtained in the calibration procedure, where the direction of a vector is given by $\beta_x$ and $\beta_y$, and the color shows the magnitude $|\bm{\beta}|$. Every other vector along each direction is shown for clarity. (b) A portion of dot pattern is shown: pink dots (tuned from black) are from a calibration image, while cyan dots (tuned from black) are from a measurement image of surface ripples, and the overlap is in blue. (c) The displacements $\delta\textbf{r}$ (amplified $4$ times in length for clarity) between the pink and the cyan dots in (b). }\label{fig:calib}
\end{figure}

The displacement $\delta \mathbf{r}(x,y)$ can be obtained by taking the cross-correlation operation over the reference and the refraction images of dot objects (see figure~\ref{fig:calib}b, c). Image cross-correlation algorithms, as commonly used in two-dimensional PIV, are applied in this study. In this study, the multi-step algorithm is used with interrogation window $96\times96$ pixel$^2$ at the initial step reducing to the window $24\times24$ pixel$^2$ with $75\%$ window overlap at the final step, an optimized setting in Lavision Davis$^\copyright$. 
The interrogation window at the final step includes approximately 8 dots for accurate displacement tracking \citep{raffel2018PIV}, given that the number of dots in each interrogation window between the paired images is approximately unchanged. In this work, the dot objects are printed (randomly distributed) on a transparent sheet. The random distribution of the dots helps the PIV correlation algorithm reduce fortuitous pattern coincidence and renders better accuracy. We carefully designed the pattern, where the diameters of individual dots are approximately 5 to 6 pixels in images. 

The camera viewing angles, $\bm{\beta}(x,y)=\beta_x(x,y)\hat{x}+\beta_y(x,y)\hat{y}$, are determined by the spatial position of the camera to the object pattern (e.g. see figure~\ref{fig:calib}a), and can easily obtained in a calibration procedure. For a leveled flat bottom and a still (flat) water surface (i.e.\ $\nabla h=0$, known $n_l$), the surface height $h_c$ can be measured with a caliper precisely. According to equation~\eqref{eq:simp}, $\bm{\beta}(x,y) = -\delta \mathbf{r}_c(x,y)/[h_c(1-n_a/n_l)]$ with $\delta \mathbf{r}_c(x,y)$ obtained from the cross-correlation operation. Note that the height $h_c$ can be freely chosen and does not depend on the details of the experiments (e.g.\ water fill level) to be performed in the tank. 
\nomenclature{$h(x, y)$}{height of surface, or height of liquid}
\nomenclature{$\nabla h(x, y)$}{Spatial gradients of surface height}
\nomenclature{$\delta \mathbf{r}(x,y)$}{Displacement vectors caused by refraction}
\nomenclature{$\delta \mathbf{r}_0(x,y)$}{Displacement vectors in calibration}
\nomenclature{$\widehat{\mathbf{n}}$}{Normal direction or orientation of liquid surface}
\nomenclature{$n$}{Refractive index}

We solve for $h(x,y)$ in equation~\eqref{eq:full} and \eqref{eq:simp} using the Matlab intrinsic function \textit{fsolve} (Levenberg–Marquardt algorithm), which requires long computing time and large memory. To reduce the need of computation power and time, an \textit{in-house} Matlab code implementing the Newton--Raphson method was scripted to solve the equation~\eqref{eq:simp}. The gradient $\nabla h$ is discretized with central, second-order finite differences at interior grid points and with forward/backward {second-order} differences at the boundaries. The displacements $\delta \textbf{r}(x,y)$ span $M \times N$ grid points, and at each grid point there are two components of $\delta \textbf{r}$ and one $h$. This leads to a linear system including $2MN$ equations and $MN$ unknowns. 
{This over-determined linear system {can be solved without imposed boundary conditions} by Newton iterations from an initial guess (a positive constant) to the converged solution $h(x,y)$. Convergence is assured if the maximum of the least-squared residual errors in the linear system drops below $10^{-5}$.}

We note that $\delta \mathbf{r}$ is measured with respect to the point $P$ in figure~\ref{fig:method}, whereas the coordinates of the intersection (surface) point for the solved $h$ differ from $P$. The coordinates of the solved $h(x,y)$ thus need to be remapped according to $(x,y)^{*}=(x,y)-h(x,y)\, \mathrm{tan}[\bm{\beta}(x,y)]$.


\section{Liquid ripples}\label{sec:ripples}
In this section, we report on experiments of surface ripples, where the spatially averaged surface height is approximately constant. 

\begin{figure}
	\centering
	\includegraphics[width=0.5\textwidth, trim={0cm 0cm 0cm 0cm}, clip]{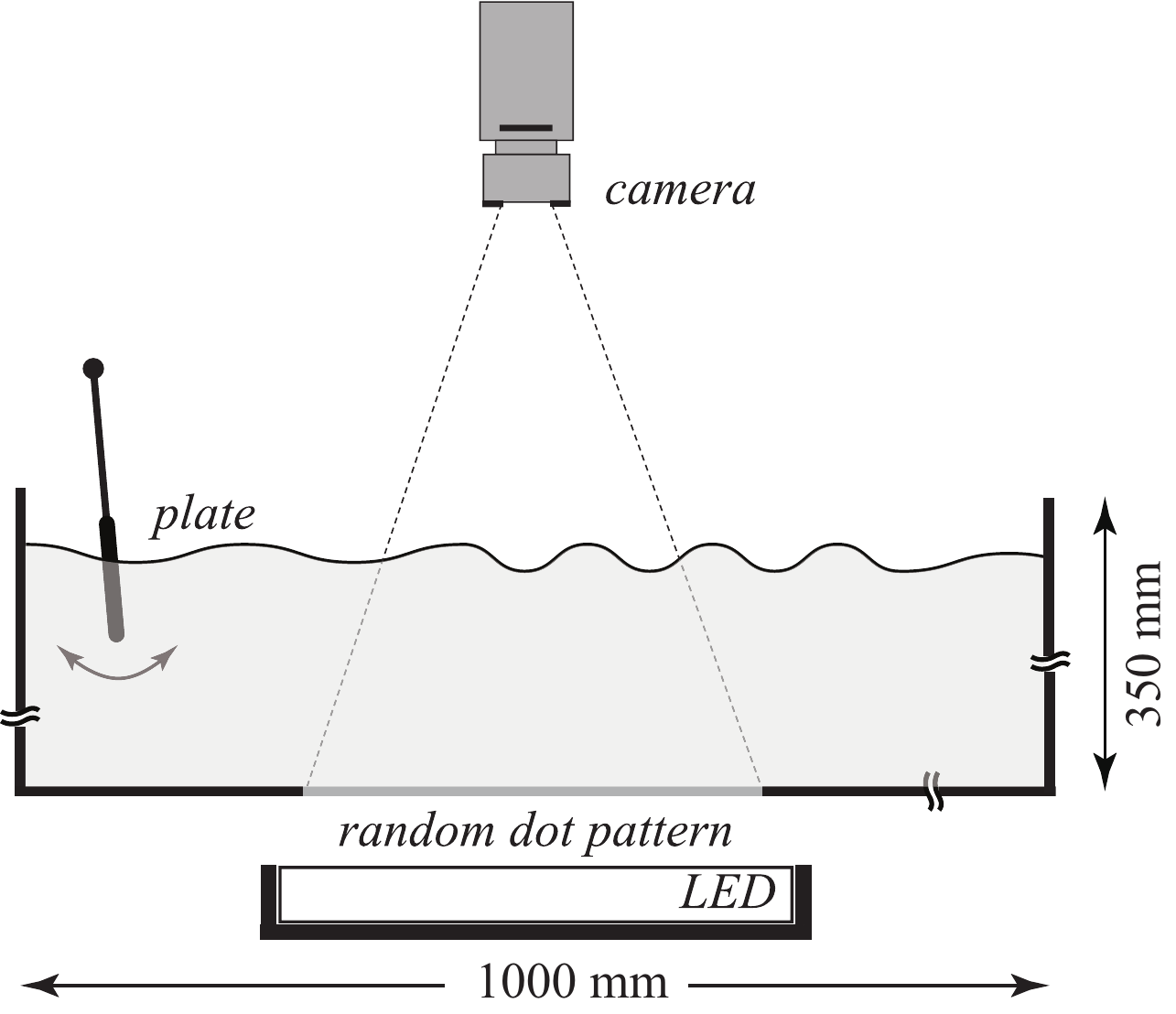}
	\caption{\label{fig:exp1} Schematic of measuring topography and height of liquid ripples in a tank. The dimensions are not in scale.
	}
\end{figure}

\subsection{Experiments}
The tank that was manufactured with transparent acrylic plates has dimensions of $1000 \times 150 \times 350$~mm$^3$ in length, width and height. The random dot pattern in dimension of $200$~mm $\times$ $120$~mm was printed on an overhead transparency (in thickness of $0.16$~mm) in a printer (Ricoh MP C3504). The dot pattern was attached upon the inner flat bottom of the tank. A white LED with a light diffuser was applied to illuminate the dot pattern from the bottom of the tank, while a camera (Phantom VEO 640L equipped with a Zeiss lens of focal length $100$~mm) in resolution 2560 $\times$ 1600 pixel$^2$ was placed approximately $1$ meter above the pattern. A plate was employed to flap back and forth to generate surface ripples (see figure~\ref{fig:exp1}). 

The experiments were carried out in the following order. The tank bottom was leveled; An image was captured as the reference image $I_0$ when there was no water in the tank; The water was filled into the tank to a height of $20\pm0.1$~mm which was measured by a caliper, and an image $I_{h_0}$ was captured when the water surface was still. When the water ripples were generated, a series of images $I(t)$ was recorded.

When the images $I(t)$ are correlated with the image $I_0$ to get $\delta \textbf{r}(x,y,t)$, the equation~\eqref{eq:full} is solved to give $h(x,y,t)$ from the present method. For comparison, when the images $I(t)$ are correlated with the image $I_{h_0}$ to get the displacements, $h(x,y,t)$ can be obtained with the FS-SS method \citep{moisy2009synthetic}. The script of FS-SS method of \citet{moisy2009synthetic} was downloaded from their website and compared with our script of the FS-SS method to confirm that we understand their method correctly. In their script, the solution of the topographic height at the bottom left point of the computation domain is set to be zero, then the summation of the solved topographic height and the reference height gives the surface height.  

\begin{figure}
	\centering
	\includegraphics[width=1\textwidth, trim={0cm 0cm 0cm 0cm}, clip]{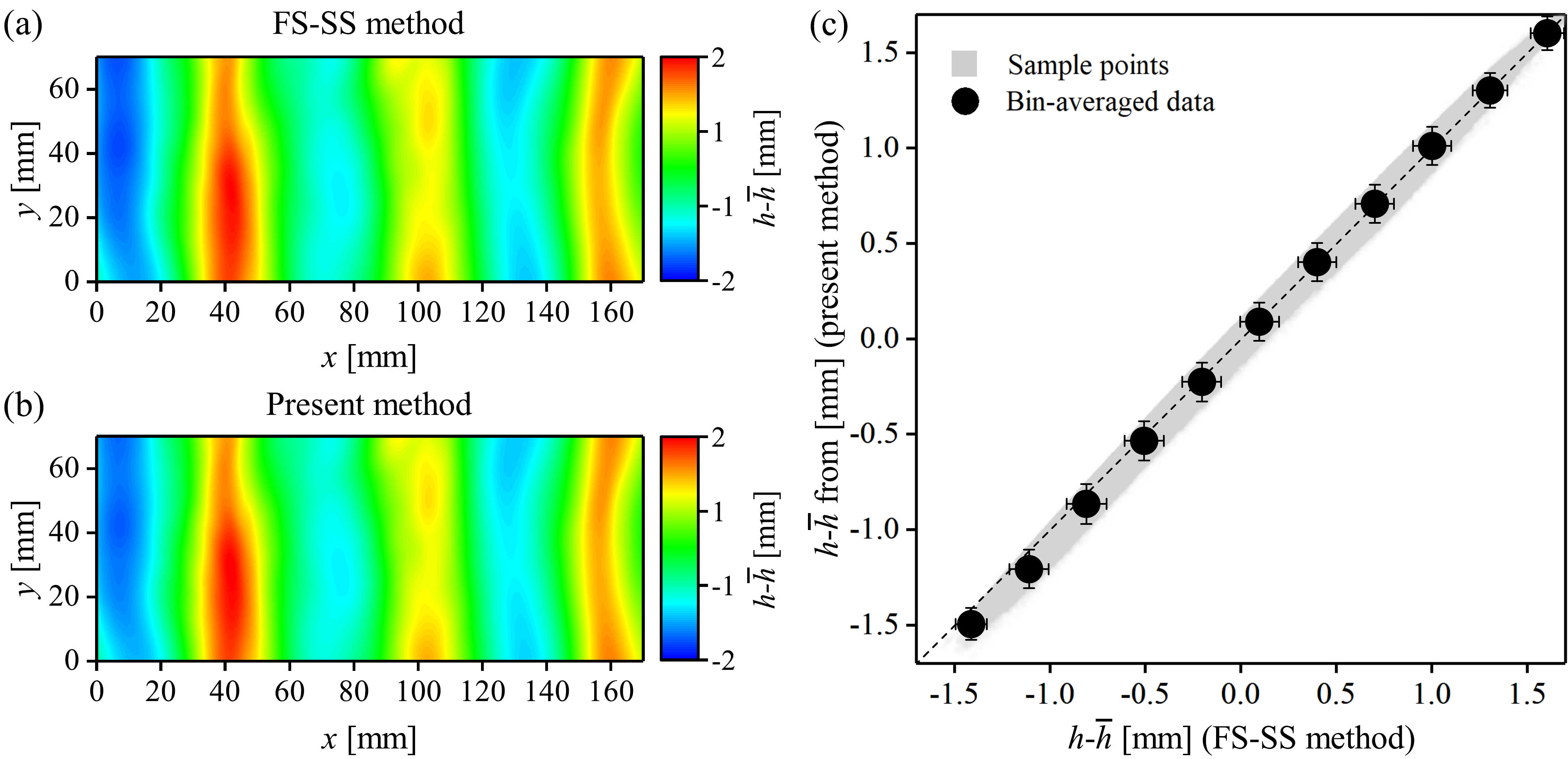}
	\caption{\label{fig:ripples} (Color online) The contours of topography from the FS-SS method (a) and from our method (b). (c) Comparison of $h-\overline{h}$ between present method and FS-SS method in experiments of dynamic ripples, where the gray squares denote individual data in space. A black dot shows the average of the individual points in a bin with the size of $0.3$~mm, and an error bar shows the standard deviation of the data in the bin. }
\end{figure}

\subsection{Results}
A snapshot of the surface ripples measured with our method is shown in figure~\ref{fig:ripples}(b), and it is visually in good agreement with the result of the FS-SS method shown in (a). 
This comparison is performed on the surface topography by removing the spatially averaged height $\overline{h}(t)$ from $h(t)$. The particular consideration behind this comparison is that, although the reference height $h_0$ is constant, the averaged height within the field-of-view measurements may vary temporally, which may break the fundamental assumption of the FS-SS method \citep{moisy2009synthetic}.
About $8,000,000$ individual spatial points of $h(x,y)-\overline{h}(x,y)$ from $200$ temporal snapshots are shown as the gray squares in figure~\ref{fig:ripples}(c), where a black dot shows the bin-averaged value of the individual points. Good quantitative agreement between the two methods is obtained.

\begin{figure}
	\centering
	\includegraphics[width=1\textwidth, trim={0cm 0cm 0cm 0cm}, clip]{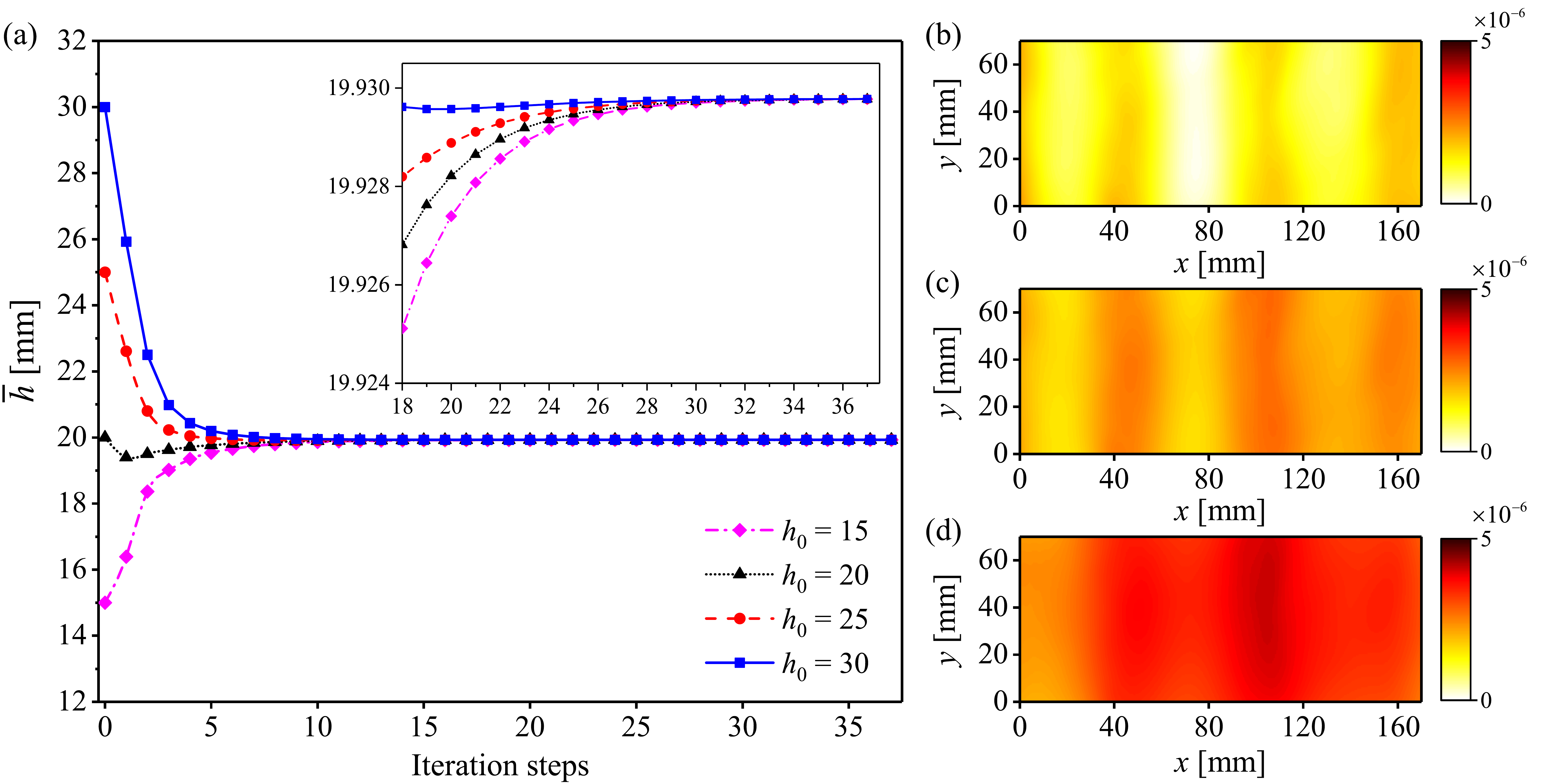}
	\caption{\label{fig:IC} (Color online) The effects of initial guesses in Newton method on $h(x,y)$ in solution. (a) The mean height $\overline{h}$ resulted from initial guesses ${h}_0 = 15, 20, 25, 30 $ against iteration steps, where the inset shows a zoom for iteration steps larger than $18$. (b--d) The contours of topography differences, $h(x,y,{h}_0 = 30)-h(x,y, {h}_0 = 20)$, $h(x,y,{h}_0 = 25)-h(x,y,{h}_0 = 20)$ and $h(x,y,{h}_0 = 15)-h(x,y,{h}_0 = 20)$, respectively.}
\end{figure}
\subsection{Effect of initial guesses}
The effect of initial guesses for the Newton-Raphson method on the computed $h(x, y)$ is shown in figure~\ref{fig:IC}(a). For the same snapshot shown in figure~\ref{fig:ripples}(b), the spatially averaged heights $\overline{h}$ computed from initial guesses ${h}_0(x,y)=15, 20, 25, 30$ converge to the same value $\overline{h}\approx19.93$ (see the inset).
The contours of the topographic difference on $h(x,y)$ between the result with ${h}_0=20$ (shown in figure~\ref{fig:ripples}b) and those with other initial guesses are shown in figure~\ref{fig:IC} (b--d), where the difference is only about $10^{-6}$. This shows that the present method does not depend on different initial guesses. When the initial guesses are meaningful, i.e. $h_0>0$, the present method does not have multi-solution issues as reported in \citet{gomit2013free}. In practice, we suggest to take $h_0(x,y)$ slightly larger than the maximum height of the surface from \textit{a-priori} knowledge of the flow.

\section{Dam-break problem: propagation of a current}\label{sec:dam_break}
A dam-break problem involves a liquid current propagating downstream with the averaged height temporally changing. The experiments of the dam-break problem introduced in this section demonstrate that with our method flows with temporally evolving averaged height can be measured.
\begin{figure}
	\centering
	\includegraphics[width=1\textwidth, trim={0cm 0cm 0cm 0cm}, clip]{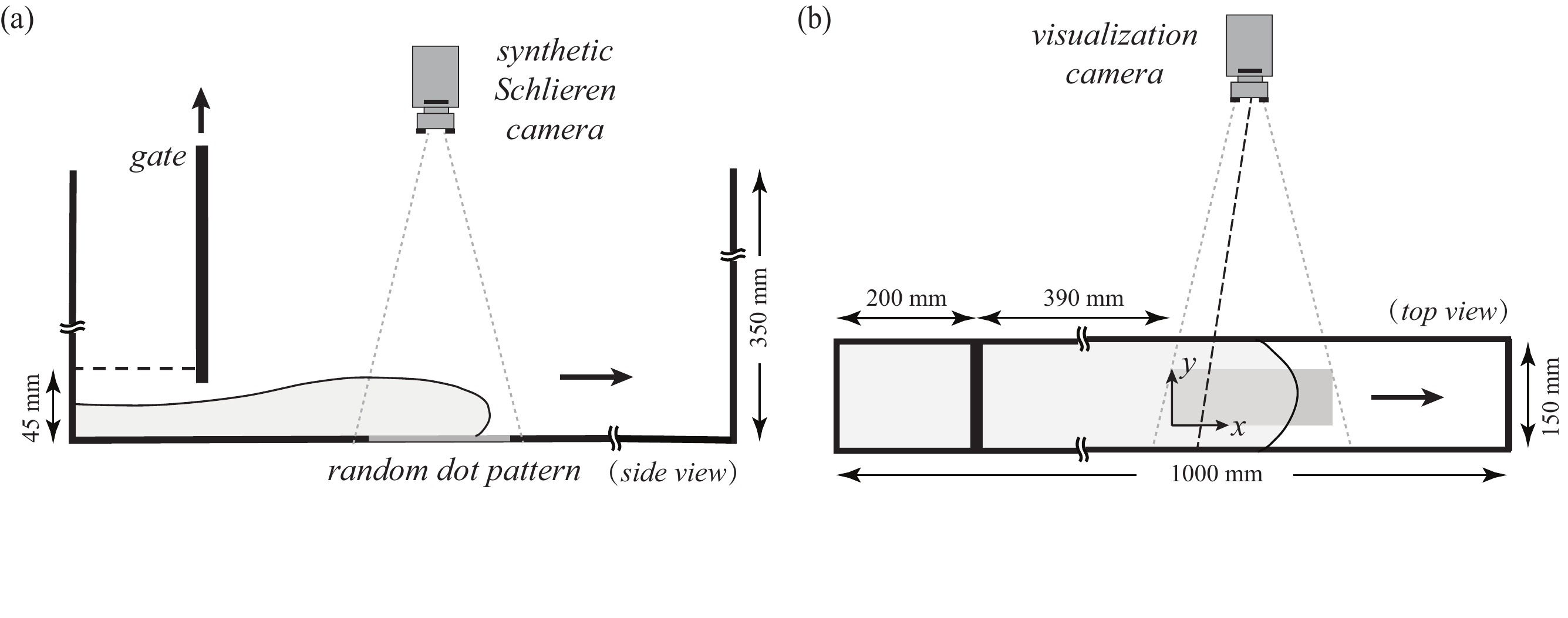}
	\caption{\label{fig:exp2} {(a) Side view schematic of a lock-exchange experimental setup for propagation of current fronts in a dam-break flow. The camera and illumination systems are the same as in figure~\ref{fig:exp1}. (b) Top-view schematic of visualization measurements. The thick black arrow indicates the flow direction, and the gray rectangular marks the area where $h(x,y)$ is measured. The dot lines enclose the field-of-view of the visualization camera, and the dash line marks the perspective view of the camera. The dimensions are not in scale.}
	}
\end{figure}

\subsection{Experiments}
The experiments of the dam-break problem were carried out in a lock-exchange setup as sketched in figure~\ref{fig:exp2}(a). The tank is the same as used in the experiments of the liquid ripples. The experimental procedure is as follows. Water was filled into the locked portion of the tank only (the left side of the gate) to the height marked by a dashed line ($45$~mm in height); The gate was rapidly, manually moved away vertically, and the water flowed into the right side of the gate driven by gravity formed a dynamical current. 

\begin{figure}
	\centering
	\includegraphics[width=1\textwidth, trim={0cm 0cm 0cm 0cm}, clip]{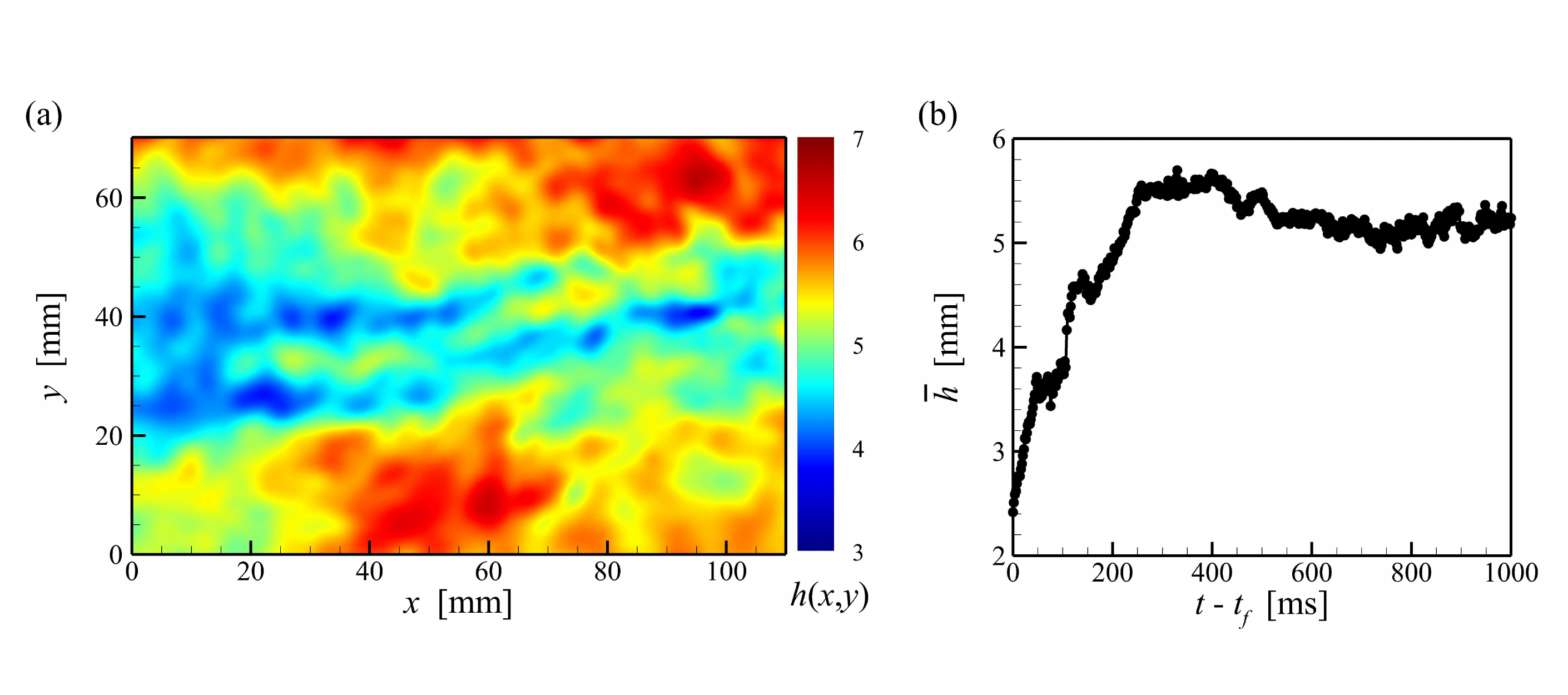}
	\caption{\label{fig:current_body} (Color online) (a) Contours of a sample $h(x,y)$ of the current body at $t-t_f=234$~ms. (b) Time series of $\bar{h}$ for the current body, where $t_f$($=152$~ms) is time duration of the current front passing the measurement domain. The gate is opened at $t\approx -40$~ms.
    }
\end{figure}

The dot pattern was attached to the inner bottom of the tank, while a camera was placed approximately $1.5$~m above the bottom (see figure~\ref{fig:exp2}a). The details of the camera-illumination system are given in \S\ref{sec:ripples}. 
An example of the current body, the surface height upstream of the current front is shown in figure \ref{fig:current_body}(a), where large spatial variations of the current heights can be observed. In time, the averaged current height $\bar{h}$ changes by $3$~mm in a short time interval ($300$~ms). As consequence, the FS-SS method is inapplicable in this flow condition.

\subsection{Measurement validation}
{For validation purposes, a visualization system was additionally set up to capture the height profile of the current at the same time. 
The visualization camera (also a Phantom VEO 640L) was placed approximately $0.7$~m away from a sidewall of the tank. A white paper was attached outside another sidewall of the tank to produce an approximately uniform background. Every liter of water was premixed with 18~ml ink to increase the contrast of the imaged current to the background. The visualization was with illumination from the same LED as the synthetic Schlieren measurement. 
This visualization measurement approximately captures the surface height averaged in the $y$ direction (see details below).
The two cameras were synchronized and the sampling rate of the measurements was $500$~frames per second. The measurement domain of the free surface was $110\times70$~{mm} in streamwise and spanwise direction, respectively. The upstream border of the measurement domain was about $390$~mm downstream of the gate. }

\begin{figure}
	\centering
	\includegraphics[width=0.95\textwidth, trim={0cm 0cm 0cm 0cm}, clip]{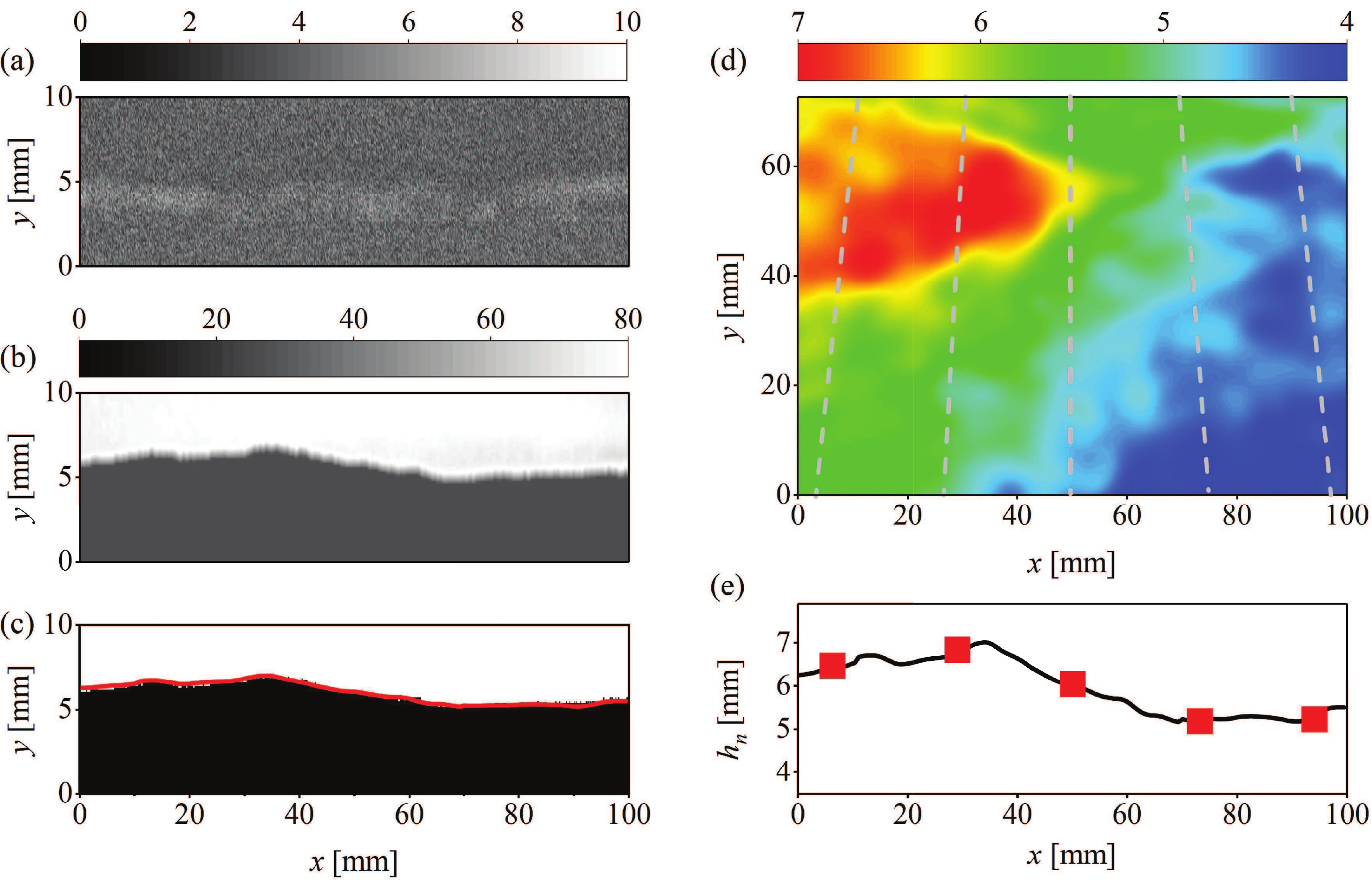}
	\caption{\label{fig:visualization} (Color online) Illustration of the image processing in the visualization measurement: (a) The image of the standard deviation of the background images; (b) A sample of visualization image; (c) The tuned image with marking the pixel with grayscale larger (or smaller) than three times of the standard deviation as black (or white), where a red line marks the height profile. Illustration of obtaining $h_n(x)$ (e) from $h(x,y)$ (d), where dashed lines indicate the perspective direction and five red squares denote the data from the five dashed lines.  
	}
\end{figure}

\begin{figure}
	\centering
	\includegraphics[width=0.9\textwidth, trim={0cm 0cm 0cm 0cm}, clip]{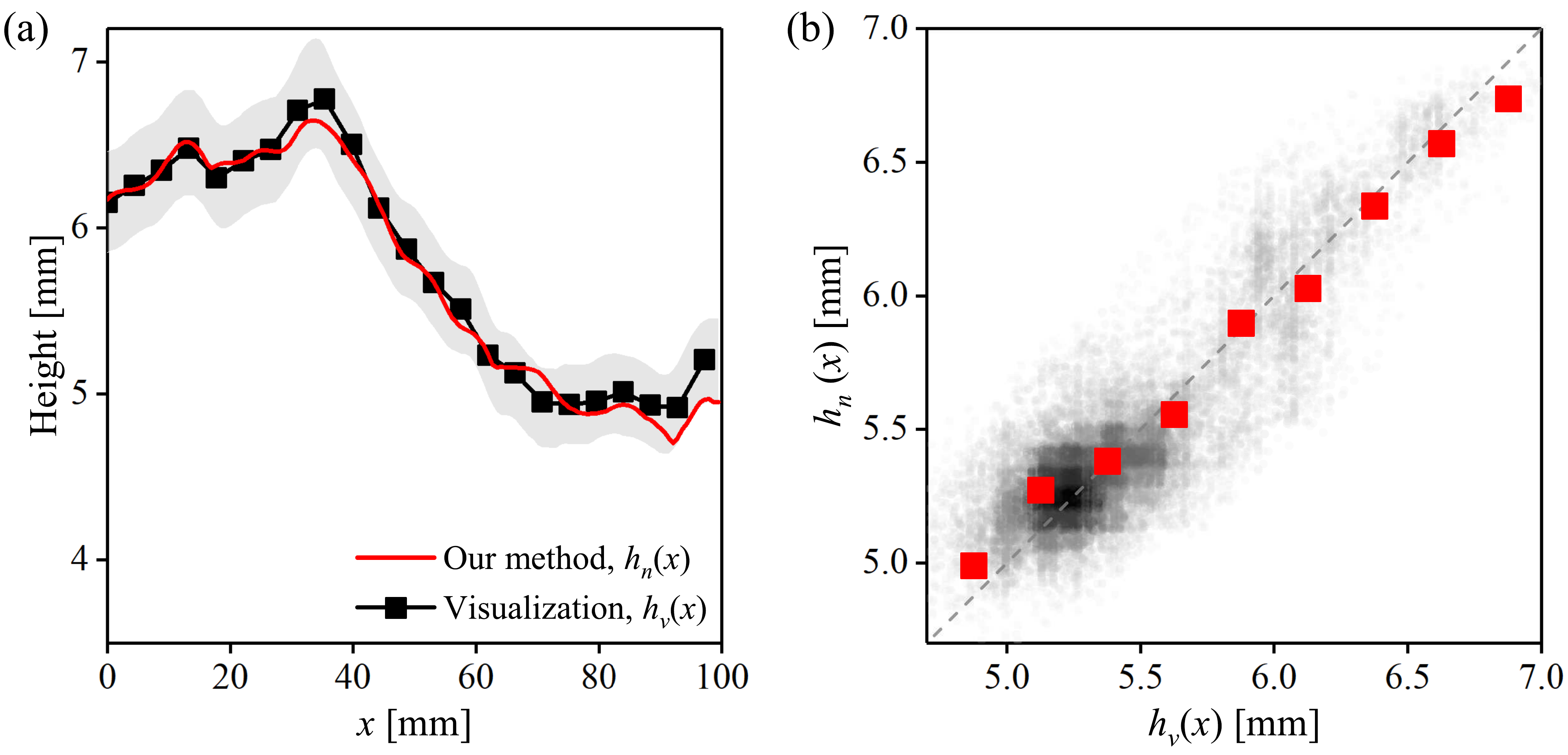}
	\caption{\label{fig:band_scatter} {(Color online) Comparison of streamwise profiles of current body between the visualization and our method. (a) The streamwise profiles $h_v(x)$ (visualization) and  $h_n(x)$ (present method) along the streamwise direction $x$. Gray band show the error bar of visualization measurement. (b) $h_n(x)$ versus $h_v(x)$, where individual data are shown in points and the grayscale level indicates the probability density function (pdf) (the darker the larger pdf), and red squares show the bin-averaged data. }
	}
\end{figure}

The height profile of the current body was obtained from the visualization measurements as follows. An image of the standard deviation (at each pixel) (see figure~\ref{fig:visualization}a) and an image of the averaged background were obtained from the background images captured before the experiments (i.e. without liquid in the measurement region). A visualization image of the current body is shown in figure~\ref{fig:visualization}(b). For the visualization image with removal of the averaged background image, if the absolute value at the pixel is larger than three times of the standard deviation at the pixel, it indicates that water flows over the pixelwise region, and the pixel is marked as black (see figure~\ref{fig:visualization}c). Similarly, if the absolute value at the pixel is smaller than three times of the standard deviation at the pixel, no water flow is expected and the pixel is marked as white (see figure~\ref{fig:visualization}c). The height profile of the current is given by the borders between the white and the black regions, then a 3-point moving average operation was used to smooth the profile to give $h_v(x)$ for the visualization measurements (see figure~\ref{fig:visualization}c). 

The visualization camera has a perspective view of the flow, and a point of the height profile approximately corresponds to the maximum height along a perspective line (see e.g. the dash line in figure~\ref{fig:exp2}b). Thus, for better comparison with the height profile of the visualization measurement, $h(x,y)$ from our synthetic Schlieren method is post-processed. $100$ points are defined with equidistance in the streamwise length of the field-of-view of the visualization camera, and they are connected to the position point of the visualization camera to define 100 perspective lines (see examples of 5 perspective lines, the dashed lines in figure~\ref{fig:visualization}d). The surface height $h(x,y)$ is interpolated along the 100 perspective lines, and the maximum value along each perspective line is obtained to give $h_n(x)$. 

We then compare $h_n(x)$ and $h_v(x)$ in figure~\ref{fig:band_scatter}(a), and they approximately collapse. About $24,000$ points of the current profiles from about $180$ snapshots are shown as gray dots in figure~\ref{fig:band_scatter}(b), where the red squares show the bin-averaged data. The two measurement results agree with small differences, which are mainly from two aspects: (1) The data of two measurements are from different sizes of the measurement domain. Although a black mask was used to block the area outside the gray rectangular in figure~\ref{fig:exp2}(b), the perspective illumination made the visualization measurement domain slightly larger than the measurement domain from the present method. (2) The visualization measurements are affected by the light reflection at the free surface and the spatially inhomogeneous illumination intensity which varies in time. Consequently, the current profiles from the visualization measurements have larger uncertainties. 

\subsection{Dynamics of current front}
\begin{figure}
	\centering
	\includegraphics[width=1\textwidth, trim={0cm 0cm 0cm 0cm}, clip]{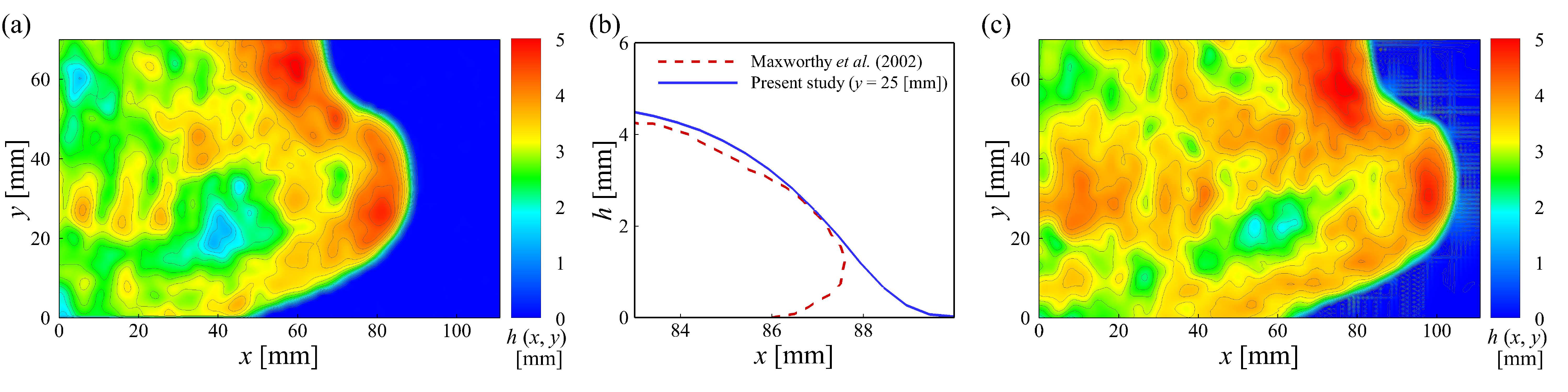}
	\caption{\label{fig:front} (a) Contours of a sample of the current front with no numerical noise. (b) The height profile of the current front, where the blue solid line shows the profile at $y=25$~mm in (a). The red dashed line shows a typical profile of a gravity current, which is extracted from the top panel of figure 4 in \citet{Maxworthy2002}, and the values are adjusted to shown in the same figure. (c) Contours of a sample of the current front with spike-like numerical noise downstream of the front. 
	}
\end{figure}
\begin{figure}
	\centering
	\includegraphics[width=1\textwidth, trim={0cm 0cm 0cm 0cm}, clip]{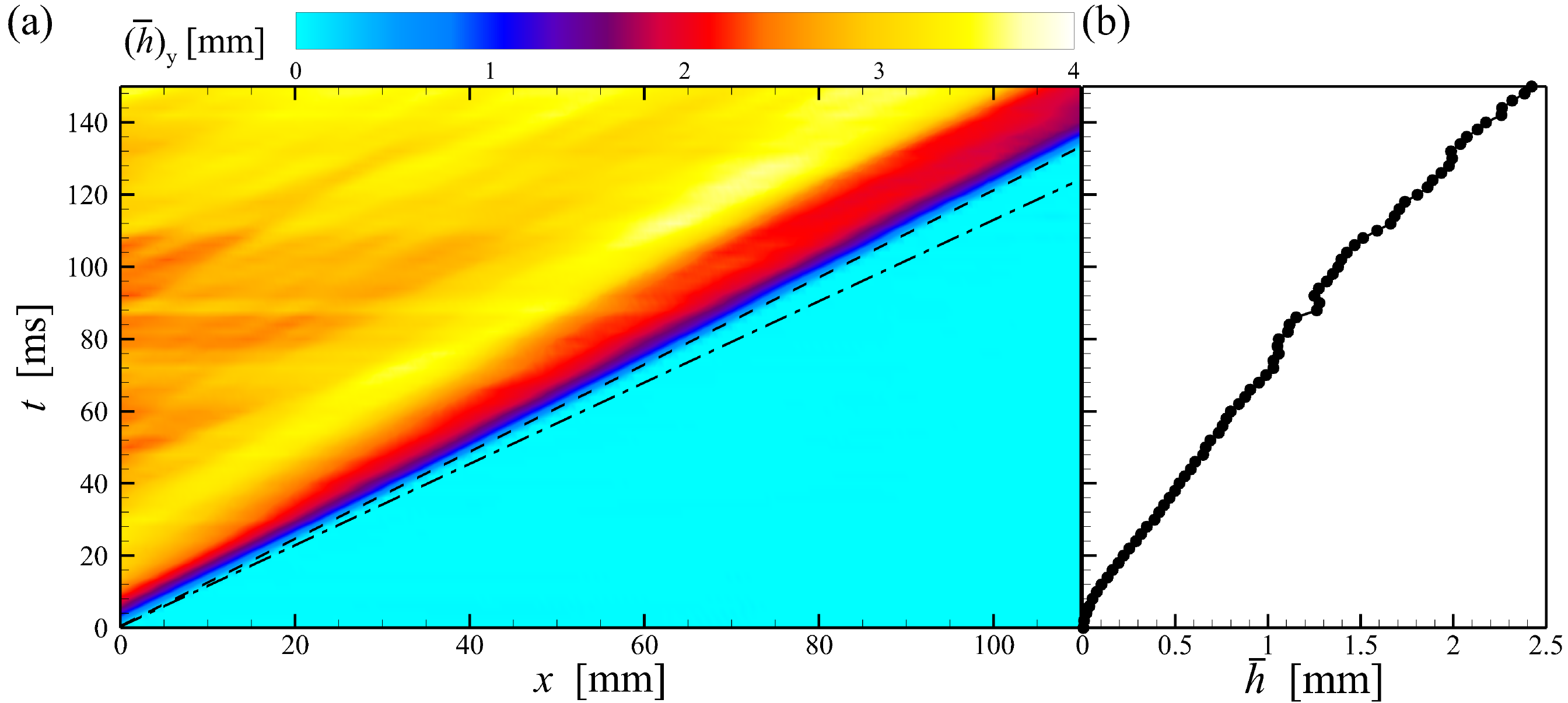}
	\caption{\label{fig:front_spacetime} (Color online) (a) Spatio-temporal diagram of the current fronts, where the color indicates the $h(x,y)$ averaged over $y$, $\bar{(h)}_y(x)=\sum_{i=1}^N h(x,y_i)/N$. The dash line indicates the front speed in our experiments. The front speed of \citet{Lowe05} is nearly the same as our experiments and is also denoted by this dash line. The dash-dot line indicates the front speed of \citet{Adegoke14}. (b) The corresponding spatial averaged height $\bar{h}(t)$, in sharing the same axis $t$.  
	}
\end{figure}
The sharp fronts where the height changes abruptly ($\nabla h \gg 1$) in a short streamwise distance (see figure~\ref{fig:front}a) results in an ill-conditioned Jacobian matrix of the Newton-Raphson method. {This leads to height errors at the downstream region of the front edge, where the height should be zero. We found that imposing a Dirichlet boundary condition at the downstream border of the computation domain (i.e. $h(\mathrm{max}(x),y)=0$) can remedy this issue. In addition, occasionally} in some snapshots, downstream of the fronts, there are spike-like structures possibly given by numerical noise (height of about $0.5$~mm, see figure~\ref{fig:front}b), the cause of which is unclear. The dynamics of the current front (after removing the noise downstream of the front) and the body can be seen in a supplemental movie.

In figure~\ref{fig:front}(a), a region of large height $h\approx 5$~mm can be seen close to the front, downstream of which the height decreases to about $3$~mm with large spatial fluctuations. This experimental scenario agrees with the simulations of \citet{Shin04,Borden13}. However, a difference on the structures of the front interface can be observed (and is expected) in figure~\ref{fig:front}(b). This difference is due to the breakdown of the working principle in this region (see the discussion in \S\ref{subsec:breakdown}).
We then show the spatio-temporal diagram of the current front in figure~\ref{fig:front_spacetime}(a). The dash line marks the position of the current front at times,
and a constant propagation speed can be seen. The front speed in our experiment is approximately $0.82$~{m/s}. This agrees well with the result of \citet{Lowe05} (about $0.83$~m/s) and \citet{Adegoke14,Lauber98} (about $0.9$~m/s). In the measurement domain, the averaged height changes in time, i.e., it increases from $0$ to approximately $2.4$~mm (see figure~\ref{fig:front_spacetime}b).  

\section{Discussion} \label{sec:discussion}
Equation~\eqref{eq:simp} has a simple form, and can be solved fast for a case using our Matlab code. This is ideal for performing the uncertainty and error analysis, as well as deriving the critical condition when the working principle is broken, thus equation~\eqref{eq:simp} is used for solving the surface height in this section. 

\subsection{Measurement uncertainty from cross-correlation methods}
The presented method relies on computing the displacements of dots of a pattern via a cross-correlation method. In the present study, the configuration of the interrogation windows of size decreasing down to $24 \times 24$ pixel$^2$ with $75\%$ overlap (a suggested setting in Lavision Davis), and a sub-pixel Gaussian interpolation was used \citep{raffel2018PIV}. The uncertainty is expected to be around $0.1$ pixel and smaller, because the random dot pattern was generated to satisfy an optimal PIV condition \citep{moisy2009synthetic}. In our problem, the uncertainty on each component of displacements of an instantaneous sample is up to about $0.05$ pixel (from Lavision Davis$^\copyright$, see \citet{Wieneke2015}).

To investigate the uncertainty of the cross-correlation method on the measured surface height, the uncertainty propagation in a linear system cannot be applied to the equation~\eqref{eq:full} (or \ref{eq:simp}). Instead, random noise was added to two components of displacement fields to evaluate the uncertainty. Two types of noises were examined: Gaussian distribution with a mean of $0.1$~pixel and a standard deviation of $0.05$~pixel, where signs ($+$ or $-$) of the values are randomly chosen; white noise (ranging from $-0.15$~pixel and $0.15$~pixel). The contaminated displacements were used to calculate the surface height $\hat{h}(x,y)$, and the resulted difference is $\hat{h}(x,y)-h(x,y)$. This procedure was repeated $100$ (Gaussian) $+ 100$ (white) times, and for each time the noise was randomly generated. 

For the data shown in figure~\ref{fig:ripples} (c), the relative uncertainty averaged over space and 100 noise samples is approximately $0.2\%$ in average with a standard deviation of $1.2\%$ for the Gaussian noise, and $0.1\%$ in average with a standard deviation of $0.9\%$ for the white noise. 

The same method of the uncertainty estimation was also applied to a displacement field, which was numerically generated in equation~\eqref{eq:simp} with $h = 0.5\: \rm{sin} (0.1 \it{x} + \rm 0.08\it{y}) + \rm{10}$ (simulating ripple stripes). For the Gaussian noise, the mean relative error is $-0.3\%$ with a standard deviation of $2\%$, while the mean relative error is nearly zero with a standard deviation of $1.4\%$ for the white noise.

\subsection{Effect of camera vibration}
The measurements using our method are sensitive to camera vibration. In our preliminary set-up, the camera vibration was particular evident and it was possibly from the cooling fan of the high-speed camera, as reported elsewhere \citep{moisy2009synthetic}, or the vibration of the supporting profiles for the camera. The vibration produced approximately $0.2$ pixel in images of a still target, and the vibration was mainly along y direction in the images. To evaluate the effect of the camera vibration on surface height, a noise taking the form $\delta \mathbf{r}_\mathrm{vibration}(x,y) = \mathrm{c_r} \cdot y + \mathrm{c_t}$ is employed to mimic rotation ($\mathrm{c_r}$) and translation ($\mathrm{c_t}$) of the camera in vibration. $\delta \mathbf{r}_\mathrm{vibration}$ is then added to the measured $\delta \mathbf{r}$ to compute $h(x,y)$. The combination of $-3\times10^{-4}\leqslant \mathrm{c_r} \leqslant 3\times10^{-4}$ and $-0.2\leqslant \mathrm{c_t} \leqslant 0.2$ (in pixel) gives vibration displacement up to $0.4$ (pixel). Taking a measurement sample of $\overline{h}\approx10$ as an example (as seen in table ~\ref{tab:vibration_mean_height} in appendix), the mean height ranges from $9.3$ to $13$, producing the relative error up to approximately $30\%$ (see table~\ref{tab:vibration_error} in appendix).

Particular care is therefore necessary to stabilize the camera against vibrations. If the vibration remains intolerable, measuring the vibration simultaneously with the surface height is necessary. For instance, the border or the corner region in the field-of-view of a camera can be used to record still objects without any light refraction, while the center region of the frame can be used to measure the marker displacements for the surface height. The shifts of the still targets can be then taken for evaluation of the camera vibration, e.g. through a two-dimensional linear interpolation (with assumed form $\delta \mathbf{r}_\mathrm{vibration}(x,y) = \mathrm{c_r} \cdot y + \mathrm{c_t}$). The  $\delta \mathbf{r}_\mathrm{vibration}(x,y)$ needs to be removed from $\delta \mathbf{r}(x,y)$ before the Newton-Raphson computation.     

\subsection{Effect of resolution}

\begin{table}
	\centering
\begin{tabular}{llll}
\toprule
{Interrogation window [pixel$^2$]} & $\mathrm{max}(h)-\mathrm{min}(h)$ [mm] & max($||\nabla h||$) & $\overline{h}$ [mm] \\ \midrule
$24 \times 24$ ($0\%$ overlap)  & {0.921} & {0.165} & 9.666  \\
$48 \times 48$ ($0\%$ overlap)  & {0.882} & {0.116} & 10.214 \\
$96 \times 96$ ($0\%$ overlap)  & {0.852} & {0.063}  & 10.189 \\ 
\bottomrule
\end{tabular}\caption{The effect of interrogation window size on the surface height and topography of the snapshot in figure~\ref{fig:ripples}(c).}\label{tab:resolution}
\end{table}

The spatial resolution of the displacement $\delta \mathbf{r}(x,y)$ plays a key role for measurements of $h(x,y)$. The dynamic range of surface topography wavelengths is determined approximately by the dynamic range of $\delta \mathbf{r}$.
To examine the effect of resolution, a pair of images was processed with the interrogation window size of $24^2$~pixels$^2$ up to $96^2$~pixels$^2$ and $0\%$ overlap was used. As shown in table ~\ref{tab:resolution}, as the resolution is decreased (increase of window size), the topographic structures are smoother and more flat, given by the decrease of $\mathrm{max}(||\nabla h||)$ and the decrease of $\mathrm{max}(h)-\mathrm{min}(h)$, respectively, and the averaged height is correspondingly changed.  

 \begin{figure}
	\centering
	\includegraphics[width=1\textwidth, trim={0cm 0cm 0cm 0cm}, clip]{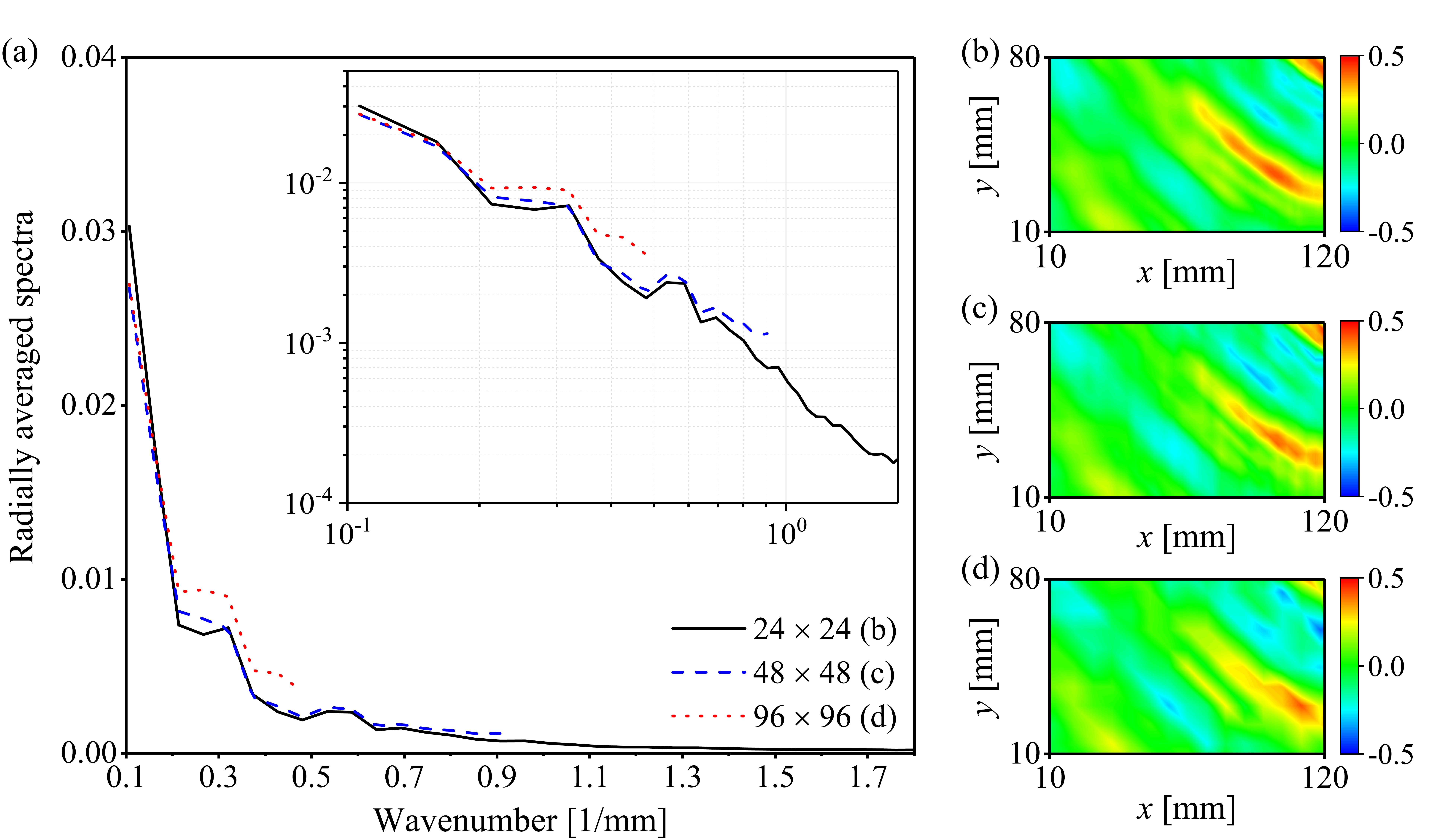}
	\caption{\label{fig:reso_fft} (Color online) (a) Radially averaged spectra of $h(x,y)-\overline{h}$ for three interrogation window sizes. The inset shows the same data in logarithm-logarithm axes. (b--d) Contours of the same region obtained with interrogation window $24^2$, $48^2$ and $96^2$~pixels$^2$, respectively.}
\end{figure} 
The two-dimensional power spectrum was calculated for $h(x,y)-\overline{h}$ to examine the fluctuations of the heights, and the radially averaged power spectrum is shown in figure~\ref{fig:reso_fft}(a). The curve of $48^2$ nearly collapses with that of $24^2$ for moderate and large topography structures (see b and c). For an interrogation window of $96^2$~pixels$^2$, the spectra curve deviates that of $24^2$, while the noticeable difference on the surface topography can be seen in (b) and (d).

\subsection{Breakdown condition of working principle}\label{subsec:breakdown}
\begin{figure}
	\centering
	\includegraphics[width=0.8\textwidth, trim={0cm 0cm 0cm 0cm}, clip]{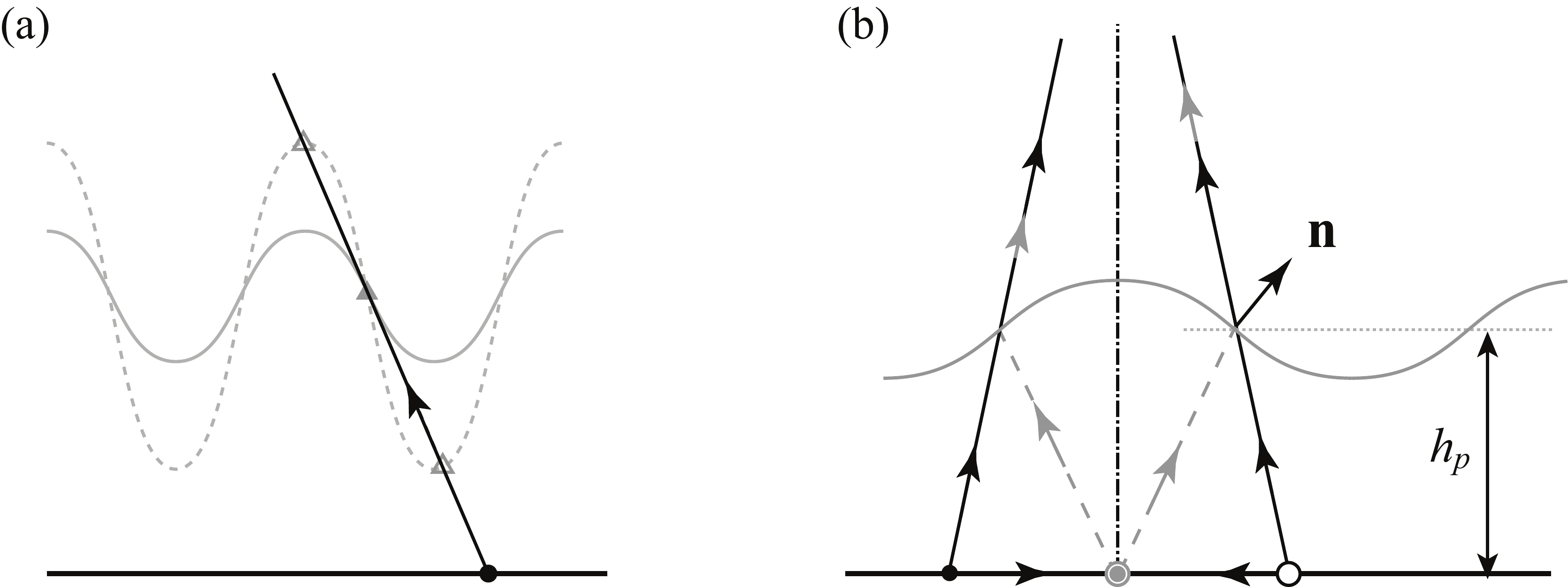}
	\caption{\label{fig:caustics} The illustration of breakdown condition of the working principle: (a) Large oscillation amplitude of surface wave, where the light ray is refracted more than once as marked by hollow triangles. (b) Critical condition for formation of caustics.}
\end{figure}
In the present method, the governing equation is physically correct provided that the light is refracted \textit{once} on the air-liquid interface. The working principle of the present method is violated when a light ray is refracted more than once or caustics is formed. The critical condition for the former case can be depicted: For a given cosinusoidal plane wave $h(x) = h_p + \eta_0 \,\mathrm{cos}(2 \pi x / \lambda)$ with wavelength $\lambda$ and amplitude $\eta_0$, the critical wave amplitude is $\eta_c=\lambda[1/4-\beta_x/(2\pi)]$. At the same wavelength, when $\eta_0>\eta_c$, light ray may be refracted more than once, see the dashed line in figure~\ref{fig:caustics}(a). For the snapshot in figure~\ref{fig:ripples}(c), $\lambda\approx60~\mathrm{(mm)}$ and $\beta_x \approx0.08~\mathrm{(rad)}$ give $\eta_c\approx15~\mathrm{(mm)}>4~\mathrm{(mm)}$, and rays are refracted only once at the surface.
The formation condition of the caustics depends on the viewing angle and the curvature of the surface \citep{gomit2013free}. Taking the same cosinusoidal plane wave as an example, substituting $h(x,y)=h_p+\eta_0\mathrm{cos}(2\pi x/\lambda)$ into the equation~\ref{eq:simp} with $\delta \mathbf{r}(\lambda/4)\approx h_p\beta_x+\lambda/4$, the critical height is $h_{p,c}=\lambda^2/[8\pi\eta_0(1-n_a/n_l)]$ taken the minimum of $h_p$. When $h_p<h_{p,c}$, no caustics is expected to be formed. However, for the very front of the current in the dam-break flow, the light rays are refracted more than once.

An \textit{a posteriori} method was introduced in \cite{moisy2009synthetic} to evaluate the formation of the caustics. When the largest extensional strain of the displacement field is smaller than one, no formation of caustics is expected. We used this \textit{a posteriori} method to examine all snapshots in this study. For instance, the maximum eigenvalue for the snapshot in figure~\ref{fig:ripples}(c) is $0.193$ ($<1$) and 0.025 ($<0.15$) \citep{moisy2009synthetic}.

\subsection{The effect of linearizing the governing equation}\label{subsec:simp_full}
\begin{figure}
	\centering
	\includegraphics[width=1\textwidth, trim={0cm 0cm 0cm 0cm}, clip]{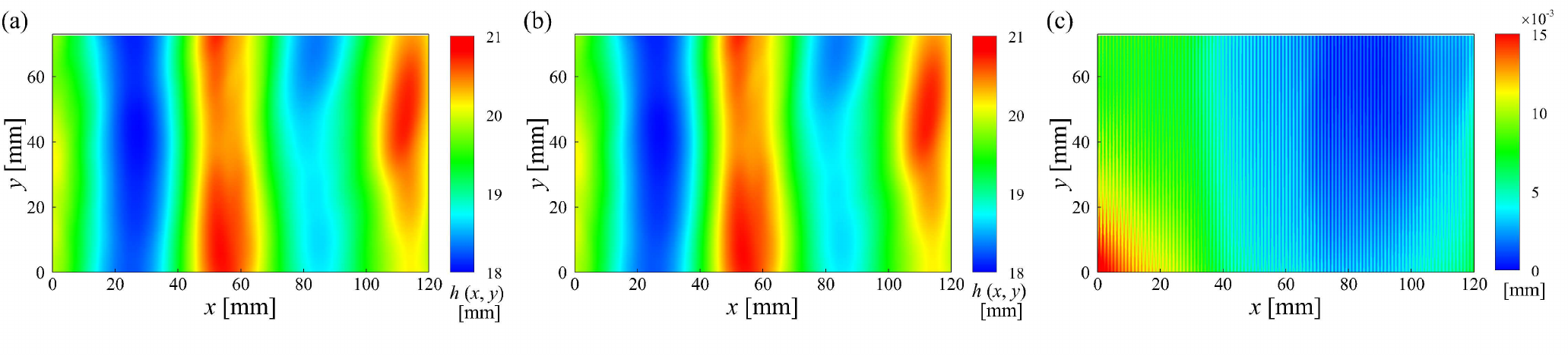}
	\caption{\label{fig:ripple_difference} (Color online) Contours of a sample of the surface ripples (the same snapshot as in figure~\ref{fig:ripples} which is obtained by solving (a) the linearized equation~\eqref{eq:simp}, (b) equation~\eqref{eq:full} and (c) their difference. }
\end{figure}
\begin{figure}
	\centering
	\includegraphics[width=1\textwidth, trim={0cm 0cm 0cm 0cm}, clip]{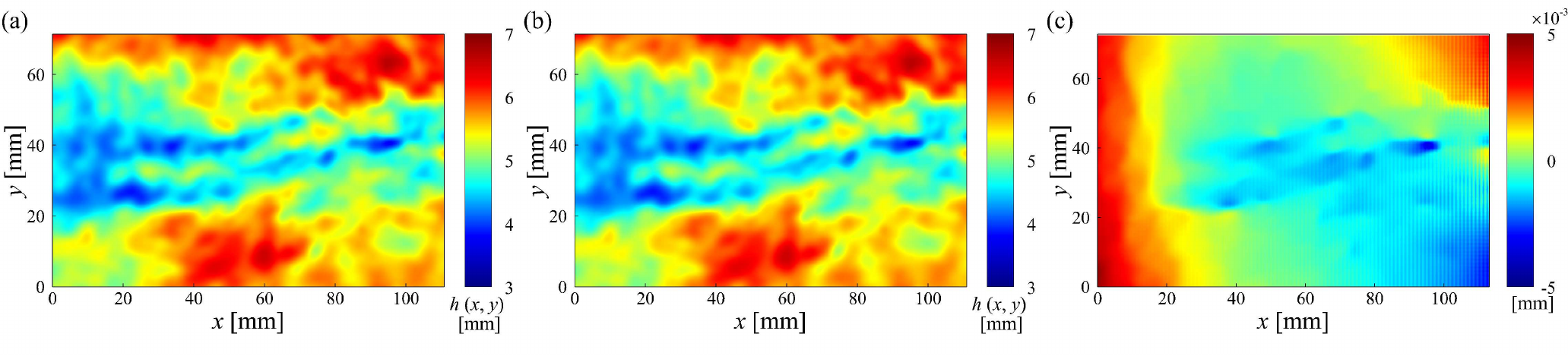}
	\caption{\label{fig:current_body_difference} (Color online) Contours of a sample of the current body (the same snapshot as in figure~\ref{fig:current_body}) obtained by solving (a) the linearized equation~\eqref{eq:simp} and (b) equation~\eqref{eq:full} and (c) their difference.}
\end{figure}
\begin{figure}
	\centering
	\includegraphics[width=1\textwidth, trim={0cm 0cm 0cm 0cm}, clip]{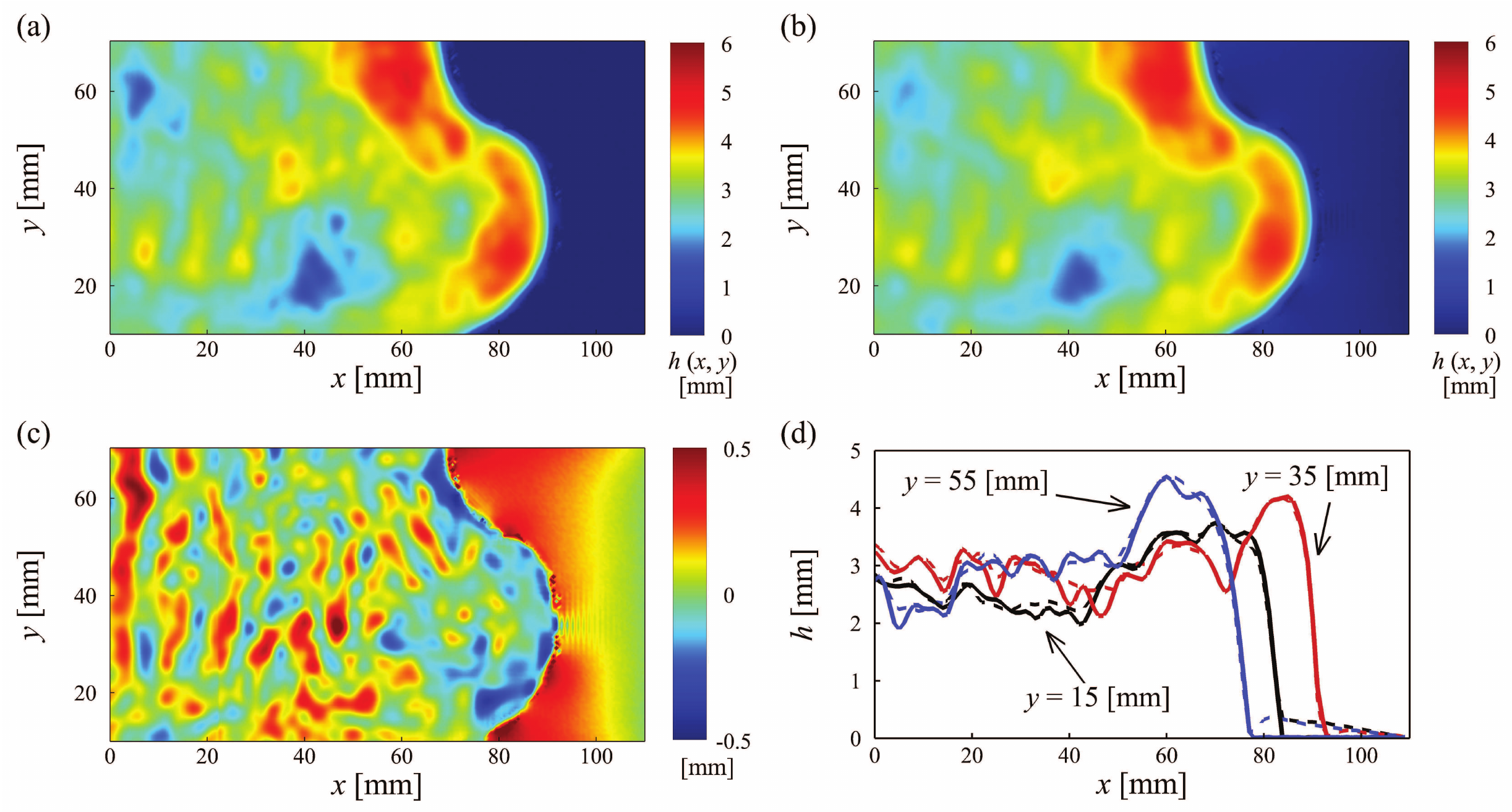}
	\caption{\label{fig:current_front_difference} (Color online) Contours of a sample of the current body (the same snapshot as in figure~\ref{fig:front}) obtained by solving (a) the linearized equation~\eqref{eq:simp} and (b) equation~\eqref{eq:full} and (c) their difference. (d) Three height profiles chosen from (a) (solid lines) and (b) (dashed lines) at $y=15$~mm, $35$~mm and $55$~mm.}
\end{figure}
{Equation~\eqref{eq:simp} is obtained by linearizing the governing equation~\eqref{eq:full}, e.g., taking $\tan{\nabla h}\approx \nabla h$ and $\tan{\pmb{\beta}}\approx \pmb{\beta}$, by truncating the second and higher order terms in their respective Taylor expansion. For example, at $\nabla h=0.5$, $\tan(\nabla h)\approx 0.5463$ and the relative difference is about $8\%$. The influence of the approximation on the solution is examined in the following.}

For the snapshot of the surface ripples shown in figure~\ref{fig:ripples}(b), the solution of the equation~\eqref{eq:simp} and the equation~\eqref{eq:full} is shown in figure~\ref{fig:ripple_difference}(a) and (b), respectively. The difference between (a) and (b) is shown in (c). The mean relative difference is about $0.02\%$, and the standard deviation is also about $0.02\%$.  
For the snapshot of the current body shown in figure~\ref{fig:current_body}(a), the solution of the equation~\eqref{eq:simp} and the equation~\eqref{eq:full} is shown in figure~\ref{fig:current_body_difference}(a) and (b), respectively. The difference between (a) and (b) is shown in (c). The mean relative difference is about $10^{-3}\%$ and the standard deviation is about $0.1\%$.  

For the snapshot of the current front shown in figure~\ref{fig:front}(a), the solution of equation~\eqref{eq:simp} and equation~\eqref{eq:full} have visual differences (see figure~\ref{fig:current_front_difference}a and b). The difference between the two solutions, as shown in (c), gives the mean relative difference about $12\%$ and the standard deviation of about $30\%$. Such difference is expected, since in the front $\nabla h\gg 1$ which violates the linearization approximation. Examination of the height profiles shows that the difference also exists downstream of the front (see figure~\ref{fig:current_front_difference}d). Although the same Dirichlet boundary condition was employed for both solvers, the noise at downstream of the front for the equation~\eqref{eq:full} remains noticeable, whereas that for the linearized equation is nearly removed. 

The linearized equation is accurate for $\nabla h<1$, whereas for $\nabla h \gg1$ (the current front), solving equation~\eqref{eq:full} is expected to give better results. However, downstream of the front there is noticeable noise albeit the Dirichlet boundary condition is imposed. In consideration that solving equation~\eqref{eq:full} takes much more computation power, the linearized equation is therefore preferred.

\section{Conclusion} \label{sec:conclusion}
In this paper, we introduce a single-camera synthetic Schlieren method to measure the topography and height of dynamic free surface. In this method, displacements of the dot patterns are attributed to both the surface height and its spatial gradients, while the reference height is not required. This method is thus applicable to flows with spatially averaged height changing in time. The dot displacements are obtained by the cross-correlation operation, while calibration is carried out to obtain the viewing angle of a camera toward the pattern object. The surface height, the only unknown in the governing equation, can be solved via a Newton-Raphson method with a proper initial guess.

We carried out two experiments to demonstrate the present method. In the experiments of surface ripples, the spatial averaged height is nearly constant. No particular treatment is performed on the boundary conditions at the borders of the computation domain. The measurement results of the present method and those of the FS-SS method of \citet{moisy2009synthetic} agree well. In another experiment, the averaged height of the current in a dam-break problem changes with time. For the current front, a Dirichlet boundary condition is applied to the downstream boundary condition of the computation domain to amend the ill-conditioned Jacobian matrix in the Newton-Raphson solver (possibly due to the sharp current front). The Dirichlet condition is not necessary for the current body, but only when the front passes the measurement section.

The presented method is still susceptible to some limitations, when the working principle can be satisfied. Specifically, the measurement results are sensitive to the camera vibrations, a problem also reported in the FS-SS method. The vibrations change the spatial relation between the camera and the dot pattern, consequently the solutions are affected particularly on the averaged surface height. Thus, particular care is required to improve the stability of the camera as well as other components of the experimental setup. When further improvement on the setup stability is impossible, a camera with a lens of long focal length is suggested while keeping the same field-of-view being viewed, the solution is found to be less sensitive to the vibrations in our practical tests. An alternative suggested solution is to take the measurements of free surface using a portion of the camera sensor, in the meanwhile use the other portion of the sensor to record the still object which is not affected by light refraction on imaging.

\section*{Appendix}\label{sec:app}

\begin{figure}
	\centering
	\includegraphics[width=0.5\textwidth, trim={0cm 0cm 0cm 0cm}, clip]{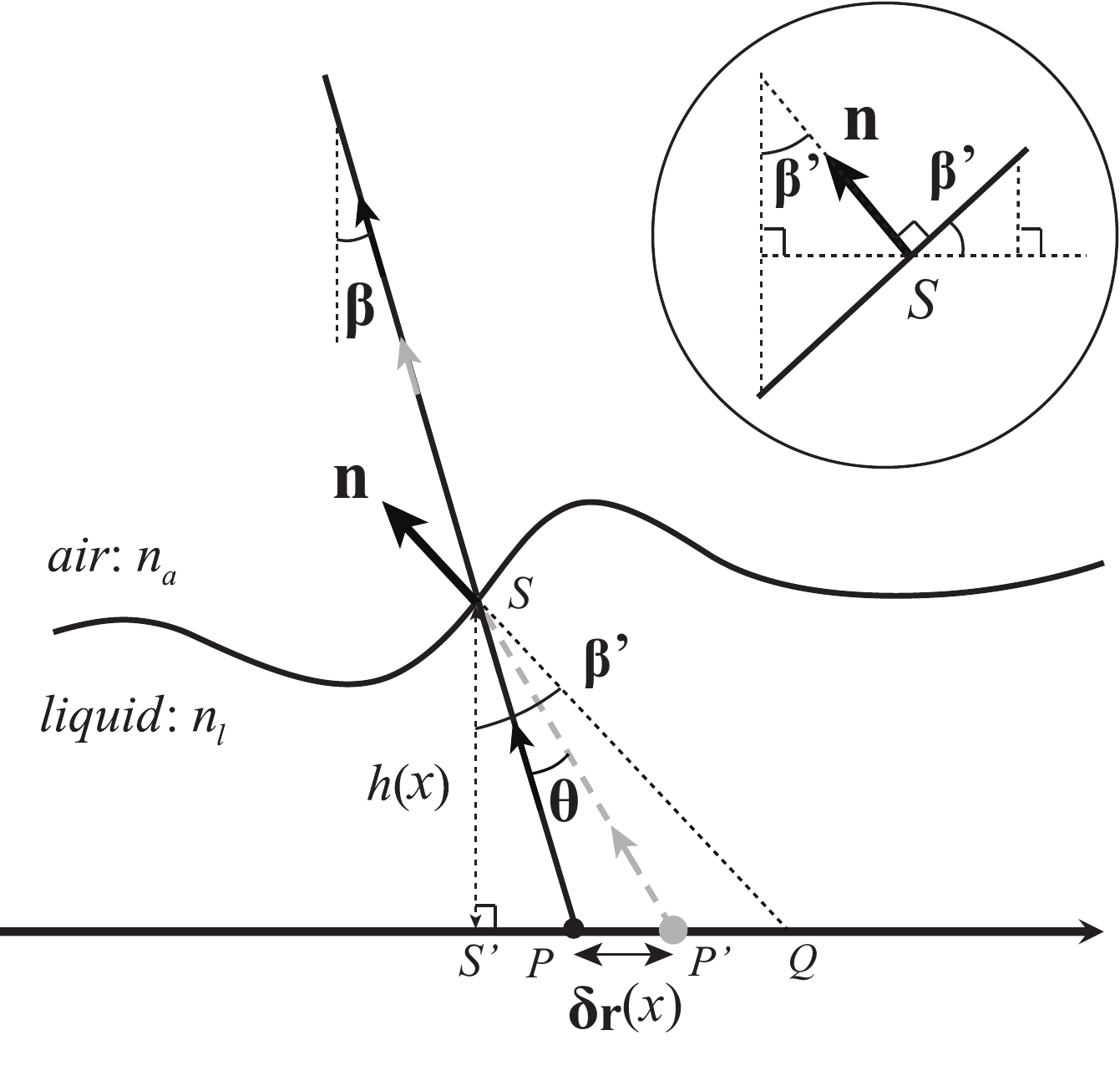}
	\caption{\label{fig:sketch_derivation} Schematic of the derivation of working principle. The inset shows the zoom-in around the point $S$.}
\end{figure}
The derivation of the governing equation~\eqref{eq:full} is shown below. From the sketch in figure~\ref{fig:sketch_derivation}, we have
\begin{equation}
	\begin{cases}
		&S'P=h\;\tan(\pmb{\beta}),	\nonumber\\ 
		&S'P'=h\;\tan(\pmb{\beta}+\pmb{\theta}),\nonumber\\
		&\delta\textbf{r}=S'P'-S'P,\nonumber\\
		&\tan(\pmb{\beta'})= \nabla h,\nonumber\\
		&\sin(\pmb{\beta}'-\pmb{\beta}-\pmb{\theta})\;n_l=\sin(\pmb{\beta}'-\pmb{\beta})\;n_a, \nonumber
\end{cases}
\end{equation} 
where the first four equations are obtained from geometry and the last one is from the Snell's law. Combining the above equations to eliminate $\pmb{\theta}$ gives the equation~\eqref{eq:full},
\begin{equation}
	\delta \textbf{r} = h \cdot \mathrm{tan}\left(\mathrm{tan^{-1}}(\nabla h) + \mathrm{sin^{-1}}[(n_a/n_l) \cdot \mathrm{sin}(\pmb{\beta}-\mathrm{tan^{-1}}(\nabla h))]\right) - h \cdot \tan(\pmb{\beta}). \nonumber
\end{equation}

\begin{table}[h]
	\centering
	\begin{tabular}{llllllll}
		\toprule
		\diagbox[width=5em]{$\mathrm{c_t}$}{$\mathrm{c_r}$} & -3 $\times 10^{-4}$ & -2 $\times 10^{-4}$  & -1$\times 10^{-4}$ & 0 & 1 $\times 10^{-4}$  & 2 $\times 10^{-4}$ & 3 $\times 10^{-4}$  \\ \midrule
		-0.20              & 10.280 & 9.917  & 9.629  & 9.480  & 9.538  & 9.817   & 10.257  \\
		-0.15              &  9.995  & 9.672  &  9.467 & 9.452  & 9.668  & 10.071   & 10.577  \\
		-0.10              & 9.739   &  9.485 &  9.401 & 9.546  & 9.902  &10.387    & 10.925   \\
		-0.05              & 9.534   & 9.387 & 9.457  & 9.754  &  10.209  &  10.737  & 11.285   \\
		0.00               & 9.406  & 9.403  & 9.632  & 10.046 &  10.557 & 11.101   & 11.647  \\
		0.05               &  9.386 & 9.542 & 9.903  & 10.388 & 10.924   & 11.469   &12.006   \\
		0.10               & 9.486  & 9.785 & 10.234 & 10.755 & 11.296  &11.835     & 12.359  \\
		0.15               &  9.697 & 10.100 & 10.597 & 11.130 &11.667  &12.193   & 12.704   \\
		0.20               & 9.988 & 10.454 & 10.972 & 11.505 &  12.031  &  12.544  &  13.040  \\ 
		\bottomrule
	\end{tabular}\caption{The averaged surface height $\overline{h}$ computed from the displacements including the camera vibration $\delta \mathbf{r}+\delta \mathbf{r}_\mathrm{vibration}$ for the snapshot shown in figure~\ref{fig:ripples}c, and $\delta \mathbf{r}_\mathrm{vibration}(x,y)=\mathrm{c_r} y + \mathrm{c_t}$. }\label{tab:vibration_mean_height}
\end{table}

\begin{table}[h]
	\centering
	\begin{tabular}{llllllllll}
		\toprule
		\diagbox[width=5em]{$\mathrm{c_t}$}{$\mathrm{c_r}$}  & -3 $\times 10^{-4}$ & -2 $\times 10^{-4}$  & -1$\times 10^{-4}$ & 0 & 1 $\times 10^{-4}$  & 2 $\times 10^{-4}$ & 3 $\times 10^{-4}$  \\ \midrule
		-0.20              & 2.33  & -1.28 & -4.15  & -5.63 & -5.06   &  -2.28   &  2.10    \\
		-0.15              & -0.51  & -3.72  & -5.77  & -5.91 & -3.76  &  0.25  &  5.29  \\
		-0.10              & -3.05 & -5.58  & -6.42  & -4.97 &  -1.43  &  3.40    & 8.75    \\
		-0.05              & -5.10  & -6.56  &  -5.86   & -2.90 & 1.62  &  6.88   &  12.34   \\
		0.00               &  -6.37 &  -6.40  &  -4.12 & 0.00  & 5.08     &10.51    & 15.94   \\
		0.05               &  -6.57  &  -5.02 &  -1.42 & 3.41  &   8.74  & 14.17    &19.51   \\
		0.10               &  -5.57 &  -2.59 & 1.88  & 7.06  & 12.45    &17.81    & 23.02   \\
		0.15               &  -3.47 & 0.54  &  5.49  & 10.79 & 16.14    &  21.38    &26.46 \\
		0.20               &  -0.57 &  4.06  & 9.22  & 14.52 &  19.76   &24.87   & 29.81   \\ \bottomrule
	\end{tabular}\caption{The relative errors (\%) of $\overline{h}(x,y)$ referring to the case $c_r=0$ and $c_t=0$ in table~\ref{tab:vibration_mean_height}.} \label{tab:vibration_error}
\end{table}

\begin{acknowledgements}
The authors thank Katja Kr{\"o}mer and Peter Prengel for their technical supports. Baofang Song is acknowledged for guiding our attention to Newton methods to solve the governing equation. Huixin Li thanks for the support from China Scholarship Council (No. 201804910530). This work was supported by the Deutsche Forschungsgemeinschaft (DFG, German Science Foundation) - INST 144/464.
\end{acknowledgements}

%
\section*{Conflict of interest}
The authors declare that they have no conflict of interest.

\bibliographystyle{spbasic}      
\bibliography{wave}

\end{document}